\documentclass[journal]{IEEEtran}

\usepackage{xkeyval}
\usepackage{multicol}
\usepackage{lipsum}
\usepackage{amsmath,amssymb,amsfonts}
\usepackage{algorithmic}
\usepackage{graphicx}
\usepackage{subcaption}
\usepackage{textcomp}
\usepackage{xcolor}
\usepackage{comment}
\usepackage{dsfont}
\usepackage{cleveref}
\usepackage{placeins}
\usepackage{bbm}
\AtBeginDocument{%
  \providecommand\BibTeX{{%
    \normalfont B\kern-0.5em{\scshape i\kern-0.25em b}\kern-0.8em\TeX}}}

\newtheorem{proposition}{Proposition}

\ifCLASSINFOpdf
\else
\fi

\allowdisplaybreaks

\begin{document}
\title{A Distributed Hierarchy Framework for Enhancing Cyber Security of Control Center Applications
}

\author{Chetan~Kumar~Kuraganti,
Bryan~Paul~Robert,
Gurunath~Gurrala,~\IEEEmembership{Senior~Member,~IEEE}, 
Ashish~Joglekar,
Arun~Babu~Puthuparambil,
Rajesh~Sundaresan,~\IEEEmembership{Senior Member,~IEEE,} Himanshu~Tyagi,~\IEEEmembership{Senior Member,~IEEE,} 
    	
    	\vspace{-1cm}
    	
    	\thanks{
    		This work was supported in part by the Bosch Research and Technology Centre, Bengaluru, India and by the Robert Bosch Centre for Cyber-Physical Systems, Indian Institute of Science, Bengaluru, India, (under Project E-Sense: Sensing and Analytics for Energy Aware Smart Campus), and in part by the Science and Engineering Research Board (grant no. EMR/2016/002503).
    	}
    	
    }


\maketitle

\begin{abstract}
Recent cyber-attacks on power grids highlight the necessity to protect the critical functionalities of a control center vital for the safe operation of a grid. Even in a distributed framework one central control center acts as a coordinator in majority of the control center architectures. Such a control center can become a prime target for cyber as well as physical attacks, and, hence, a single point failure can lead to complete loss of visibility of the power grid. If the control center which runs the critical functions in a distributed computing environment can be randomly chosen between the available control centers in a secure framework, the ability of the attacker in causing a single point failure can be reduced to a great extent. To achieve this, a novel distributed hierarchy based framework to secure critical functions is proposed in this paper. The proposed framework ensures that the data aggregation and the critical functions are carried out at a random location, and incorporates security features such as attestation and trust management to detect compromised agents. A theoretical result is proved on the evolution and convergence of the trust values in the proposed trust management protocol. It is also shown that the system is nominally robust so long as the number of compromised nodes is strictly less than one-half of the nodes minus 1. For demonstration, a Kalman filter-based state estimation using phasor measurements is used as the critical function to be secured. The proposed framework's implementation feasibility is tested on a physical hardware cluster of Parallella boards. The framework is also validated using simulations on the IEEE 118 bus system. 
\end{abstract}


\IEEEpeerreviewmaketitle
\vspace{-0.5cm}
\section{Introduction}
\IEEEPARstart{P}{ower} grids are becoming potential targets for various kinds of cyber and physical attacks because of the massive economic and social disruptions that can arise in the event of a widespread and prolonged outage of the electric grid. A comprehensive survey of data attacks against power system operations and control is presented in \cite{Bas2020Arxiv}. For example, an attacker may gain operational access to the SCADA control center and could disrupt the power grid's operation \cite{Nguyen2020}. Different detection, protection, and mitigation strategies are presented in \cite{Nguyen2020} to enhance the resilience and operational endurance of the energy delivery infrastructure against cyber attacks. Various research directions to evaluate the cyber security and develop novel algorithms for securing future power state estimation and grid operation are presented in \cite{Cha2019Springer}.

In recent years, there have been significant research efforts in securing SCADA and Energy Management Systems (SCADA-EMS) \cite{Lakshminarayana2020, Nguyen2020}, which are the backbone of energy control center operations. A class of false data injection attacks (FDIAs) on contingency analysis (CA) through state estimation (SE) is proposed in \cite{Kan2018IEEE, Yua2018IEEE}. A secure distributed state estimation method is proposed in \cite{Cin2019IEEE} to address false data injection threat in distributed microgrids. A residual detector method to detect individual compromised sensors is proposed in \cite{Bas2019ACC}. A reinforcement learning-based FDIA attack strategy is presented in \cite{Liu2019IEEE}. 
A class of sparse undetectable attacks (SUAs), designed to deteriorate the state estimation performance, is presented in \cite{Yan2020SciD}. In \cite{Cha2020IEEE}, a robust method to detect random errors and cyber attacks targeting AC dynamic state estimation through false data injection is presented.

Distributed processing is widely used in large power systems \cite{sha_2004_JWS} where a central control center coordinates the activities of sub-area control centers. Distributed processing of load flow, state estimation, security assessment, etc., requires a coordinator to first gather the sub-area data, and then process them to get the entire system view. This coordinator is usually fixed in the existing installations. Even though processing is distributed and loss of one sub-area data can be handled by the coordinator, the loss of the coordinator itself leads to the loss of visibility and control of the entire system. To avoid this, back-up control centers are put in place. Even so, simultaneous attacks on the coordinator and its backup control center can bring the system down. If the role of the coordinating control center is randomly switched among the existing control centers and automatic detection of compromised control centers, their isolation, and subsequent recovery are enabled, then the possibility of complete loss of system visibility and controllability can be reduced. In particular, single point of failure vulnerability is mitigated to a great extent. With this motivation, a distributed hierarchy-based framework is proposed in this paper. 
Note that in a centralized environment, intelligent electronic devices (IEDs) are responsible for data exchange between the control centers. 
Future IEDs \cite{samie2019edge} will have the capability to perform not only state estimation, but also complex data analysis and many more of the control center's system level algorithms in a fully decentralized manner. In such scenarios, one or more IEDs in a sub-area may act as coordinators for that area's activity, visibility, and control. 


From a broader perspective, this work enables a distributed implementation of the meta-function of monitoring the integrity of critical functionalities, with state estimation in a power system being one such critical function example. We focus on the integration of various existing schemes to prove the proposed concept using a physical demonstration. While development of new security features is beyond the scope of this paper, we provide a novel theoretical analysis of a security feature that suggests significant nominal robustness for scenarios where up to nearly half of the number of agents may be compromised (more precisely, up to strictly less than $N/2 - 1$). 
The hierarchy is obtained through a leader election process. Attestation and  trust management schemes are mainly used to detect malicious devices.

\vspace{-0.3cm}
\section{Proposed Distributed Hierarchy Framework}
The proposed distributed hierarchy framework assumes the presence of a proprietor, who is responsible for coordinating each agent and running critical algorithms in the network. This includes the code/programs/firmware that the agents will run. An agent can represent an IED having edge compute capabilities or a computer in a control center responsible for coordination and running various algorithms. An attacker may compromise some agents by corrupting some of the processes associated with these agents. Specifically, the attacker corrupts some part of the code running on the compromised agent. As a consequence, the compromised device actions may not follow the proprietor’s protocol. Such compromised devices are termed malicious. Data from these agents may no longer be reliable and if any of the analysis algorithms executed by the proprietor have dependency on this data, they may result in inaccurate results. In this paper, SE is the critical algorithm that is to be secured. A completely connected network is assumed between agents, and the devices exchange information through broadcasting. In the case of a mesh network, broadcast can be achieved via a multicast protocol that achieves broadcast. In this framework all the agents are assumed to have equal capabilities in terms of computational and communication features, so that any agent can run the proprietor functions at any time. A simple leader election (LE) protocol is proposed to randomize the location of the proprietor among the devices. When there are no malicious agents in the network, SE will run normally at a randomly elected proprietor or \textit{leader} agent. If malicious agents are present, this should be detected, such agents should be isolated, and actions must be taken to continue SE. If the leader itself is a malicious agent, SE must continue through another agent automatically without interruption.

\begin{figure}[!htb]
	\vspace{-0.4cm}
	\centering
	\includegraphics[width=\linewidth,height=1.5in]{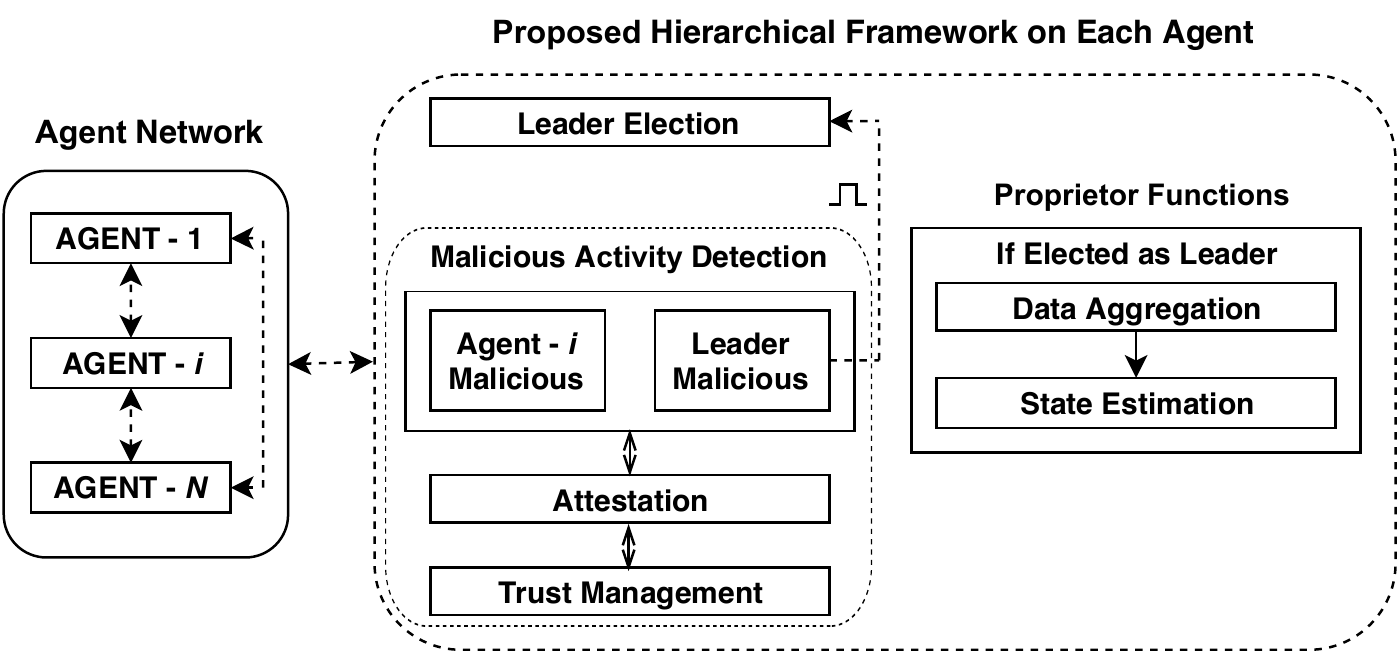}
	\caption{Overview of the distributed hierarchy framework.}
	\label{Fig:over_scheme}
	\vspace{-0.7cm}
\end{figure}

An overview of the proposed framework is shown in Fig.\ref{Fig:over_scheme}. The functional modules of the proposed hierarchical framework, shown in the figure, are:\\
\textit{Leader Election:} This module hosts leader election algorithm.
\textit{Proprietor Functions:} This module aggregates the data from other agents and runs the system level algorithms (e.g. state estimation).\\ 
\textit{Malicious Activity Detection:}  This module checks the integrity of the code running on the agents using attestation mechanisms. Further, using a trust management scheme this module infers the presence of malicious agents in the network and identifies whether a leader is malicious or any other agents are malicious.

Initially, agents communicate with each other to establish a network. Then, a leader election scheme is employed to transfer the control to a single randomly elected agent, where the SE will be performed. In parallel, attestation and trust management are continuously used to monitor the network for compromised agents, if any, and leader will ignore all the data from these agents. If the leader itself is identified to be compromised, say by a majority of the agents, a trigger to elect a new leader from among the non-malicious agents is issued. A new leader is then elected and control is subsequently handed over to it. 
\begin{figure*}[!htb]
	\centering
	\includegraphics[width=\linewidth,height=1in]{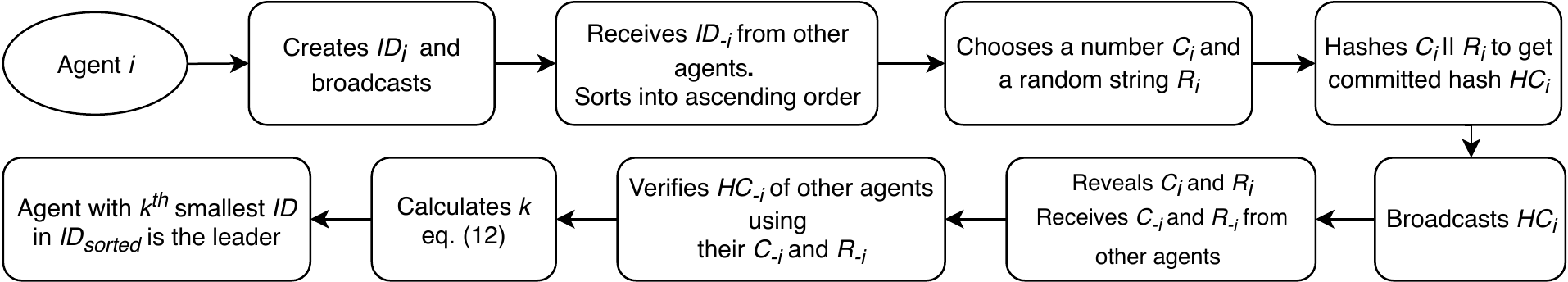}
	\caption{Overview of the leader election scheme.}
	\label{fig:le_scheme}
	\vspace{-0.7cm}
\end{figure*}
\subsection{Leader Election}\label{subsection:LE} Leader election is a fundamental problem in distributed computing. A single agent, among the many in the network, is chosen to coordinate and perform a critical meta-function, SE in this paper. The election of the leader is done in a distributed way. The proposed scheme, inspired by \cite{abraham2013distributed}, not only tries to prevent malicious agents from hijacking the election process, but also ensures that each agent has an equal probability of being elected. Numerous protocols are available for this purpose (see \cite{Lyn_1997_acm,Pet_1982_acm}) when there are no malicious agents, but \cite{fel_1997_siam,kat_2009} contain extensions to deal with malicious agents. The proposed choice of leader election scheme makes significant usage of commitment schemes \cite{damgard2008commitment}. In general, a commitment scheme follows a two stage process:
\begin{itemize}
	\item Stage 1 (Commit):  The sender locks a message in a ``box'', and sends the box to the receiver.
	\item Stage 2 (Reveal): Later, the sender provides a ``key'' to open the box, which contains the original message.
\end{itemize}
This classic cryptographic primitive was proposed in \cite{Blu_1983_acm}; see also \cite{bra_1986_ieee,gol_1988_siam}. Commitment means an agent chooses a value from a finite set and electronically commits to that choice so that it is difficult to change the value later even if it wanted to. Multiple malicious agents are assumed, even during leader election, and the malicious agents may try to influence the outcome of election. For example, one of them may desire to be the leader, or avoid becoming the leader, or favor another malicious agent to become the leader. Furthermore, during the election process agents can communicate with each other only via broadcast, and unicast/multicast modes are blocked. Availability of secure broadcast channel is a fair assumption in this scenario and the broadcast messages are taken to be \emph{common knowledge}. A detailed description of the leader election scheme shown in Fig.\ref{fig:le_scheme}, is provided below. Suppose there are $N$ agents in the network.
\begin{enumerate}
    \item Each agent \textit{i} chooses a 32-bit identity number, $ID_i$, which is broadcast to all agents. Once an agent receives $ID := [ID_0, ID_1,\dots, ID_{N-1}]$, it will then sort them in ascending order, and store them in $ID_{sorted}$. 
    \item In the next step, each agent chooses a number $C_i \in \{0,...,N-1\}$ and a random string \(R_i\).
    \item Agent $i$ then commits $C_i$. To commit $C_i$, the agent uses a cryptographic hash function $\mathcal{H}$, which takes the hash of $C_i$ appended with the random string to give $HC_i := \mathcal{H}(C_i||R_i)$.
    \item Each agent $i$ broadcasts its own hash $HC_i$, and aggregates the hashes $HC := [HC_0, HC_1,..., HC_{N-1}]$ of all agents.
    \item Agents reveal their $C_i$ and $R_i$ to every other agent (via broadcast). This information, along with respective hashes, is used to verify that the agent is committed to its $C_i$, and has not changed it.
    \item Each agent will compute $k= \Bigg(\sum_{i=0}^{N-1} C_i \Bigg)\bmod N.$.
    \item The agent with the $k^{th}$ smallest $ID$ in $ID_{sorted}$ is chosen as the leader.
\end{enumerate}
Since each agent will possess $ID_i$ from every agent $i$, $ID_{sorted}$ will be identical across agents. Similarly, all agents will be able to arrive at the same $k$, since each agent will possess $C_i$ from every agent $i$. They can thus agree on a particular agent becoming the leader without need for a central entity. 

The leader election process is triggered by the other agents when an existing leader is compromised. Then new $ID$s are generated by the agents. This forces a different ordering of agents in $ID_{sorted}$ each time. Since the collection of $C$s is used by agents to obtain $k$, a malicious agent must not be able to change its $C_i$ after learning the $C$s from other agents. Therefore commitment schemes are used to force an agent to pledge to its chosen $C_i$. In general, commitment schemes have two properties, hiding and binding. The hiding property makes it difficult for other agents to determine $C_i$ of agent $i$ from the hash, before the revelation phase. The binding property makes it difficult for agent $i$ to change its $C_i$ after commitment. The random string $R_i$ is appended to $C_i$ when hashing to randomize the resulting hash.

The algorithm above is a modification of the algorithm $A-LEAD^{ps, uni}$ in \cite{abraham2013distributed} which is used to elect a leader in an asynchronous unidirectional ring network of agents. 
The presented algorithm differs from $A-LEAD^{ps, uni}$ in two ways. First, a hash-based commitment scheme is used instead of Naor's protocol \cite{abraham2013distributed}; second, $A-LEAD^{ps, uni}$ algorithm embeds a unidirectional ring into a completely connected network. However, this approach is time consuming, since any communication must pass through the entire ring. Instead, we assume the availability of a secure broadcast channel for communication. More advanced leader election schemes such as \cite{Les2019ACM} also can be used in the proposed framework.

A hash-based commitment scheme is used, as it is practical and easy to implement. Moreover, commitment schemes built on cryptographic hash functions are known to be secure \cite{damgard2008commitment}.
The hash function in this work is taken from the libhydrogen cryptographic library \cite{jedisct1}, which is based on the Gimli permutation \cite{bernstein2017gimli}. The hash function can be replaced by other well-accepted hash functions like BLAKE and SHA3, keeping in mind the pitfalls pointed out in \cite{halevi1996practical}. 
\vspace{-0.6cm}
\subsection{Threat Model}\label{subsection:TM} \vspace{-0.2cm}
The threat model assumes a cyber-attack in which an attacker can capture an agent or group of agents, re-program them with malicious code, and then re-deploy them back into network with the intent of affecting SE. The threat model makes the following assumptions:
\begin{itemize}
    \item The number of malicious agents is $< \frac{N}{2}-1$.
    \item A compromised agent stays compromised until it is detected.
\end{itemize}
Note that physical attacks are not considered, such as transmission blockage, modification of physical sensors, enlarging the agents' memory, increasing processor speed, etc., which are other ways by which an agent's integrity could be compromised.   

\vspace{-0.5cm}
\subsection{Malicious Activity Detection}
In this section, the attestation scheme and the trust management scheme are described. Together they help detect malicious activity of agents in the network.
\subsubsection{Attestation}\label{subsection:attestation}
There are many mechanisms for detecting compromised agents, for e.g., watch-dog \cite{Mar_2000_acm}, reputation-based systems \cite{Gan_2008_acm}, etc. However, these methods are error prone, because they rely on accurate observation and reasoning of agents' misbehaviour. In this work, a distributed software-based attestation scheme, inspired by the one proposed in \cite{yang2007distributed}, to verify the integrity of the code running within an agent, without physically accessing it. Typically, software-based remote attestation uses a challenge-response protocol between two agents, as described in SWATT \cite{ses_2004_ieee}. The challenge-response protocol involves a verifier and an interrogated device (attester). The \textit{verifier} sends a challenge to the attester, which performs a checksum on its process memory, and acknowledges back with a response. Then, the \textit{verifier} validates the response, by comparing it with an anticipated response, which it has pre-computed locally.   

The proposed attestation framework differs from the others in using the concept of a \textit{report}, i.e., after every attestation, the verifier broadcasts the details of attestation to all other agents. We choose to broadcast reports to other agents to prevent tampering of the report itself. This report contains various parameters used in attestation, along with the outcome, i.e., whether the verifier suspects the attester to be malicious or otherwise. Agents make use of the reports to arrive at a consensus on whether or not malicious agents are present, and if yes, identify these malicious agents \cite{Robert2019}. The attestation algorithm is designed to check if the program code has been corrupted by an attacker. The program memory, which contains the vital code is checked during the attestation. This vital code also include the code of the proposed security algorithm. Every agent serves as a verifier at least once in a window of approximately $T$ seconds. During this interval, an agent will provision itself as a verifier at a random instant of time, and challenge a randomly chosen attester, and then stay idle for the remainder of the time interval. However, in this interval, an agent can receive multiple challenges. This process is repeated every $T$ seconds, ad infinitum. The frequency of attestations an agent can perform (the value of $T$) can be limited based on its computational capability. Details of the attestation algorithm used are not included due to lack of space; see \cite{Robert2019}. 

%

\textit{On the choice of software-based attestation}: 
Software-based attestation is an attractive choice for assessing integrity as it requires no extra hardware. However, there are attacks on software-based attestation; see \cite{kovah2012new}, \cite{castelluccia2009difficulty}. 
Use of time restrictions on the attestation process provides extra guarantees. Some attestation schemes \cite{kil2009remote}, \cite{tan2011tpm} make use of hardware components like Trusted Platform Modules (TPMs) to establish a root of trust, i.e., ability to store secret keys, calculate hashes of memory during boot, etc. Also, trusted execution environments like ARM TrustZone \cite{arm2009security} have been explored for attestation in \cite{abera2016c}, as they provide features like memory protection and isolated execution of code. Generally, hardware-based attestation techniques have better security guarantees and a wider range of use-cases than software-based techniques \cite{abera2016things}. Since the primary focus was the demonstration of a critical meta-function SE through distributed attestation and trust management, the simplest software-based attestation is chosen on account of its straightforward implementation. Any well designed attestation protocol can be integrated with the proposed scheme, for e.g. \cite{abera2016c},\cite{kil2009remote}, with suitable modification for operation in a distributed fashion can be incorporated. Indeed, the Parallella board contains ARM TrustZone that can be used to improve the attestation scheme; see \cite{Robert2019}.   

\subsubsection{Trust Management}\label{subsection:TMS}
Suppose there are $N$ agents in the network. Whenever an agent $k$ attests an agent $j \neq k$ and broadcasts its report, the evolution of trust of agent $j$ at agent $i$ at time $t+1$ can be expressed as
\vspace{-0.2cm}
\begin{equation}
\label{eqn:trust_model}
    p_{ij}(t+1) = 
    \begin{cases}
    \Big[p_{ij}(t) + \Delta_{k\rightarrow j}(p_{ik}(t))\Big]_0^1\quad & i \neq j \\
    1 & i = j,
    \end{cases}
    \vspace{-0.2cm}
\end{equation}
where $[x]_0^1$ is the projection of $x$ on the set $[0,1]$, $p_{ik}(t)$ is the trust of agent $k$ at agent $i$ at time $t$, and $\Delta_{k\rightarrow j}(p_{ik}(t))$ is (in the proposed design)
\vspace{-0.3cm}
\begin{equation}\label{eqn:indicator}
    \Delta_{k\rightarrow j}(p_{ik}(t)) = 
     \begin{cases}
       \frac{p_{ik}(t)}{N}, &\text{if attestation is positive,}\\
       -\frac{p_{ik}(t)}{N}, &\text{if attestation is negative,}\\
     \end{cases}
     \vspace{-0.2cm}
\end{equation}
where the subscript $k\rightarrow j$ indicates that agent $k$ is the verifier and agent $j$ is the attester. When the agent $j$ fails an attestation, it is called as a negative attestation, whereas if an attestation is successful, it is called as a positive attestation. It is assumed that every agent has its own opinion of trust for every other agent in the network and must perform an attestation on another agent in a set time interval. Moreover, it is also assumed that there will be no report losses during the attestation process, i.e., every report sent by an agent is received by the other agents. Handling report losses is beyond the scope of the work. At the beginning, an agent's opinion of trust for every agent in the network is initialized to $1$. 

As stated earlier, the purpose of attestation is to check the integrity of each agent by using a challenge-response mechanism. Whenever an agent $k$ attests agent $j$, agent $k$  broadcasts the status to all other agents. Then, every agent $i$, $i \in \{1,\dots, N\}\setminus\{j\}$, in the network, updates the trust of agent $j$ using \cref{eqn:trust_model} by considering its own opinion of trust of agent $j$ and agent $k$, which are $p_{ij}(t)$ and $p_{ik}(t)$, respectively. If the opinion of trust of a particular agent reduces to 0, then agents perform majority voting to cast out the malicious agent from the network. Any agent can initiate this procedure, and analogous to leader election, this too can be done in a distributed fashion. Trust of a non-malicious agent may reduce because of false accusations by the malicious agents. However, other non-malicious agents increase the trust for that agent with every positive attestation. The factor $\frac{1}{N}$ is used to ensure graceful updates.

A short summary of the analysis of a tractable variation of the proposed trust management scheme with a diminishing step size rule is provided. The result highlights interesting issues which will be discussed after stating the analytical result. Consider the following: A uniformly randomly selected agent $k$ verifies, from among others, a uniformly randomly selected attesting agent $j$; the increments in \eqref{eqn:trust_model} are modified from \eqref{eqn:indicator} to be
\vspace{-0.5cm}
\begin{equation}\label{eqn:indicator2}
    \Delta_{k\rightarrow j}(p_{ik}(t)) = 
     \begin{cases}
       a(t) p_{ik}(t), &\text{if the attestation is positive,}\\
       -a(t) p_{ik}(t), &\text{if the attestation is negative,}\\
     \end{cases}
\end{equation}
where $a(t) = 1/(t+1)$. (If $a(t) \equiv \frac{1}{N}$, we can recover \eqref{eqn:indicator}.) Now \eqref{eqn:trust_model} and \eqref{eqn:indicator2} constitute a {\em projected stochastic approximation scheme} with a step size that decreases with time.

Let $\mathcal{H}$ be the set of honest (non-malicious) agents, with the remaining agents taken to be compromised or malicious. Let 
\vspace{-0.2cm}
\begin{equation}\label{eq:defs}
  e_j  :=  \begin{cases} 
          1 & j \in \mathcal{H} \\
         -1 & j \notin \mathcal{H}.
         \end{cases} \\
\end{equation}
The quantity $p(t):= (p_{ij}(t), i \in \mathcal{H}, 1 \leq j \leq N)$ is the trust values held by each of the honest agents about each of the other agents at time $t$. Further, $p_{iH}^{(j)}(t):=  \sum_{k \in \mathcal{H}\setminus\{i,j\}} p_{ik}(t),$ and $p_{iM}^{(j)}(t):=  \sum_{k: k\notin \mathcal{H}\cup\{i,j\}} p_{ik}(t)$, are the sum of the trust values of all the honest agents except $j$ and $i$, as held by $i$ at time $t$ and the sum of the trust values of all the malicious agents including $j$ and $i$, as held by $i$ at time $t$, respectively. For $p = (p_{ij}, i \in \mathcal{H}, 1 \leq j \leq N)$, define 
$h(p) := ( h_{ij}(p), i \in \mathcal{H}, 1 \leq j \leq N ),$
where, with $[\cdot]_+$ and $[\cdot]_-$ denoting positive and negative parts, respectively,
\vspace{-0.3cm}
\begin{equation}\label{eqn:drivingfunction}
  h_{ij}(p) := \begin{cases}
     \frac{e_j \left(p_{iH}^{(j)} - p_{iM}^{(j)}\right)}{N(N-1)} & p_{ij} \in (0,1) \\
     -\frac{\left[e_j \left(p_{iH}^{(j)} - p_{iM}^{(j)}\right)\right]_{-}}{N(N-1)} & p_{ij} = 1 \\
     \frac{\left[e_j \left(p_{iH}^{(j)} - p_{iM}^{(j)}\right)\right]_{+}}{N(N-1} & p_{ij} = 0.
  \end{cases}
\end{equation}
Let $h_E$ be an enlargement of $h$ as follows: it is the smallest upper semi-continuous set-valued map with compact convex values such that $h(p) \in h_E(p)$ for almost all $p \in [0,1]^{\mathcal{H} \otimes N}$.For our security analysis, we make the following simplifying assumptions:
\begin{enumerate}
\item The attestation primitive is ideal and reveals the correct labeling $H$ and $M$ for honest and malicious agents, respectively.

\item All malicious agents cooperate to report a failed attestation on honest attesting agents and a successful attestation on other malicious attesting agents.
\end{enumerate}
Under these assumptions, we get the next result.
\begin{proposition}\label{prop:main}
Under the aforementioned threat model and attestation model, almost surely, $p(t)$ converges as $t \rightarrow \infty$ to a connected, closed, internally chain transitive invariant set\footnote{Connectedness and closedness of sets are familiar notions. Invariant sets are those that are invariant for the dynamics given by the differential inclusion $\dot{p}(s) \in h_E(p(s))$. Not all invariant sets are settlement sets for the iterations. Internally chain transitive invariant sets are specific invariant sets for the dynamics $\dot{p}(s) \in h_E(p(s))$, see \cite{borkar2009stochastic} for a definition, and the iterates settle in one such set (possibly random).} of the differential inclusion $\dot{p}(s) \in h_E(p(s))$.
\end{proposition}
\begin{IEEEproof}
See Appendix \ref{appendix:proof-of-proposition-main}.
\end{IEEEproof}
{\em Remarks}: The reason that the differential inclusion $\dot{p}(s) \in h_E(p(s))$ is tracked, and not the differential equation $\dot{p}(s) = h(p(s))$, is because $h(\cdot)$ has discontinuities on the facets of the unit cube, a fact that can be easily checked. The main reason for this is the projection operation. When the driving function for the differential equation has discontinuities, the differential equation may not be {\em well-posed}, i.e., a solution may not exist or there may be multiple solutions. One must therefore view the solution to the differential equation in a generalized, the so-called Filippov, sense; again, see \cite[Sec.~5.4]{borkar2009stochastic}. This generalization involves $h_E(\cdot)$, the smallest upper semi-continuous, compact, convex, {\em set-valued} enlargement of $h(\cdot)$, and solutions to the differential inclusion $\dot{p}(s) \in h_E(p(s))$. 

To get some insight on the consequences of Proposition \ref{prop:main}, the trust values $p^* := (p^*_{ij} = \mathbbm{1}_{\mathcal{H}}(j), i \in \mathcal{H}, 1\leq j \leq N)$, where $\mathbbm{1}_{\mathcal{H}}$ is the indicator function, is a fixed point for the dynamics $\dot{p}(s) = h(p(s))$ since $h(p^*) = {\bf 0}$. Indeed, if the agents begin the process with an initial state $p(0) = (p_{ij}(0) = 1, 1\leq i,j \leq N)$, i.e., all agents are assigned a trust value of 1, and $|\mathcal{H}| > N/2+1$, i.e., the number of malicious agents is strictly less than $N/2-1$, then $p^*$ is the point to which the differential equation dynamics will settle. 

However, there are also other fixed points for the dynamics, for e.g., $q = (q_{ij} = \mathbbm{1}_{\mathcal{H}^c}(j), i \in \mathcal{H}, 1\leq j \leq N)$. If the dynamics settles at this point, then each honest agent views every other malicious agent with a trust value of 1 but every other honest agent with a trust value of 0. This is exactly the opposite of what we desire, and we should avoid entering the basin of this attractor.

Let us now discuss the stochastic iterates. Our analysis assumes that an agent is randomly chosen to be a verifier and that this agent randomly chooses another agent for attestation. Thanks to Proposition \ref{prop:main}, the stochastic iterates track the differential inclusion dynamics. When we start at $p(0)$, with all trust values 1, we are in the setting of the first observation above, and the stochastic iterates will converge to $p^*$ with high probability, see \cite{borkar2009stochastic} for theoretical estimates of the so-called {\em lock-in} probability. However, randomness can also push the system from the initial trusting state of $p(0)$ into the basin of the undesired attractor $q$, for example when the randomness overwhelmingly chooses the malicious agents as verifiers in the verification procedure. Of course this happens with lower probability when there are a larger number of honest agents, but since this is in-principle possible, Proposition \ref{prop:main} cannot be strengthened any further. However, larger the number of honest agents, lower is the probability of randomness overwhelmingly picking the malicious agents as verifiers, and lower is the probability of settling in such undesired equilibria. In our empirical experiments, we did not encounter settlement at such undesirable equilibria; see section \ref{sec:MMA}. A way to pro-actively address the issue is to ensure that every agent verifies only once in one round of approximately $T$ seconds; results of the corresponding implementation are in section \ref{sec:MMA}.

\vspace*{-.1in}

\section{Experimental Setup and Results}\label{section:ES}
\begin{figure}[!htb]
	\vspace{-0.5cm}
	\begin{subfigure}[t]{0.5\linewidth}
		\centering 
		\captionsetup{justification=centering}
		\includegraphics[width=\linewidth, height=1.1in]{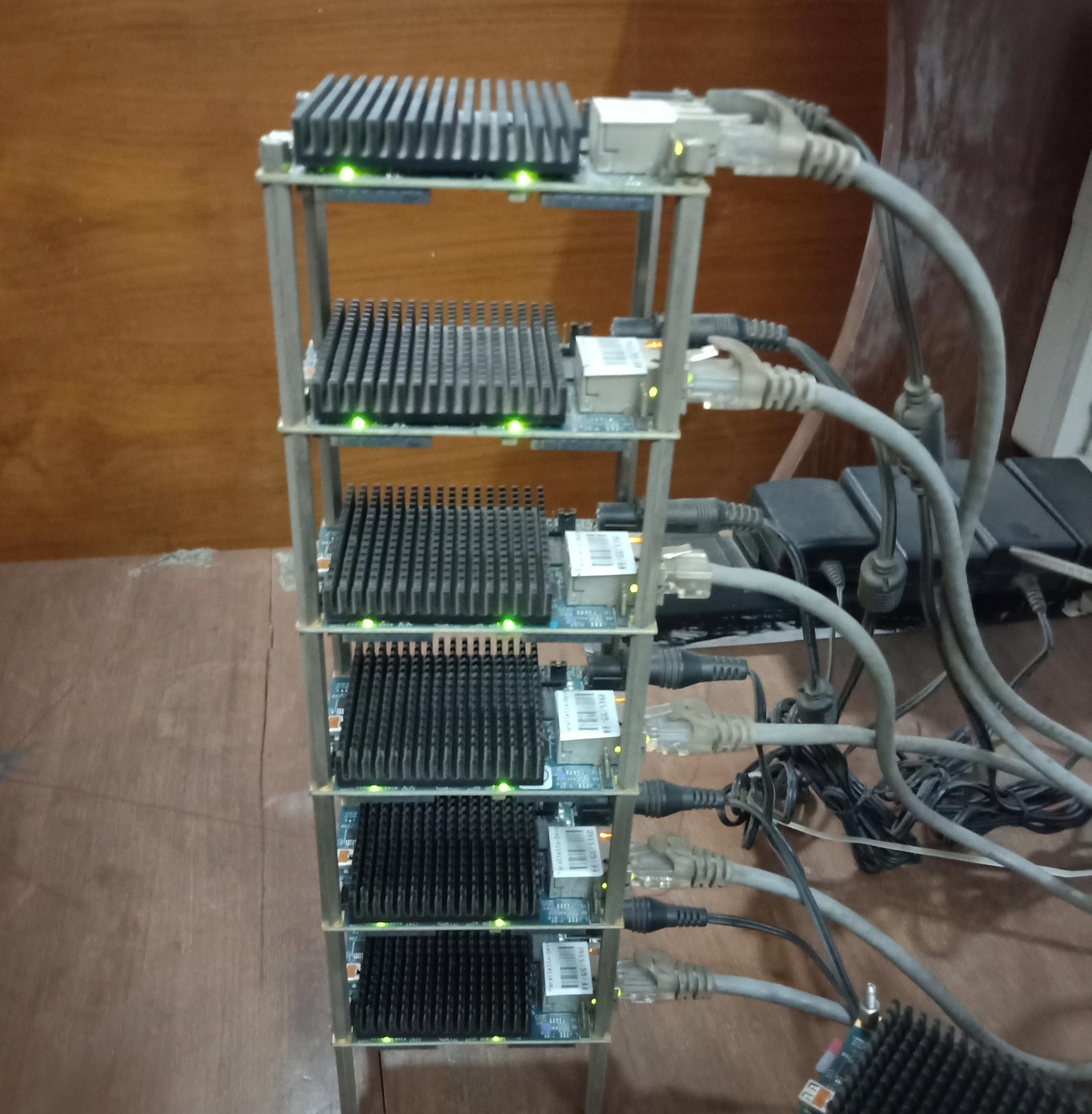}
		\caption{IED Cluster} 
		\label{fig:Parallellas}
	\end{subfigure}%
	\begin{subfigure}[t]{0.5\linewidth}
		\centering 
		\includegraphics[width=\linewidth, height=1.2in]{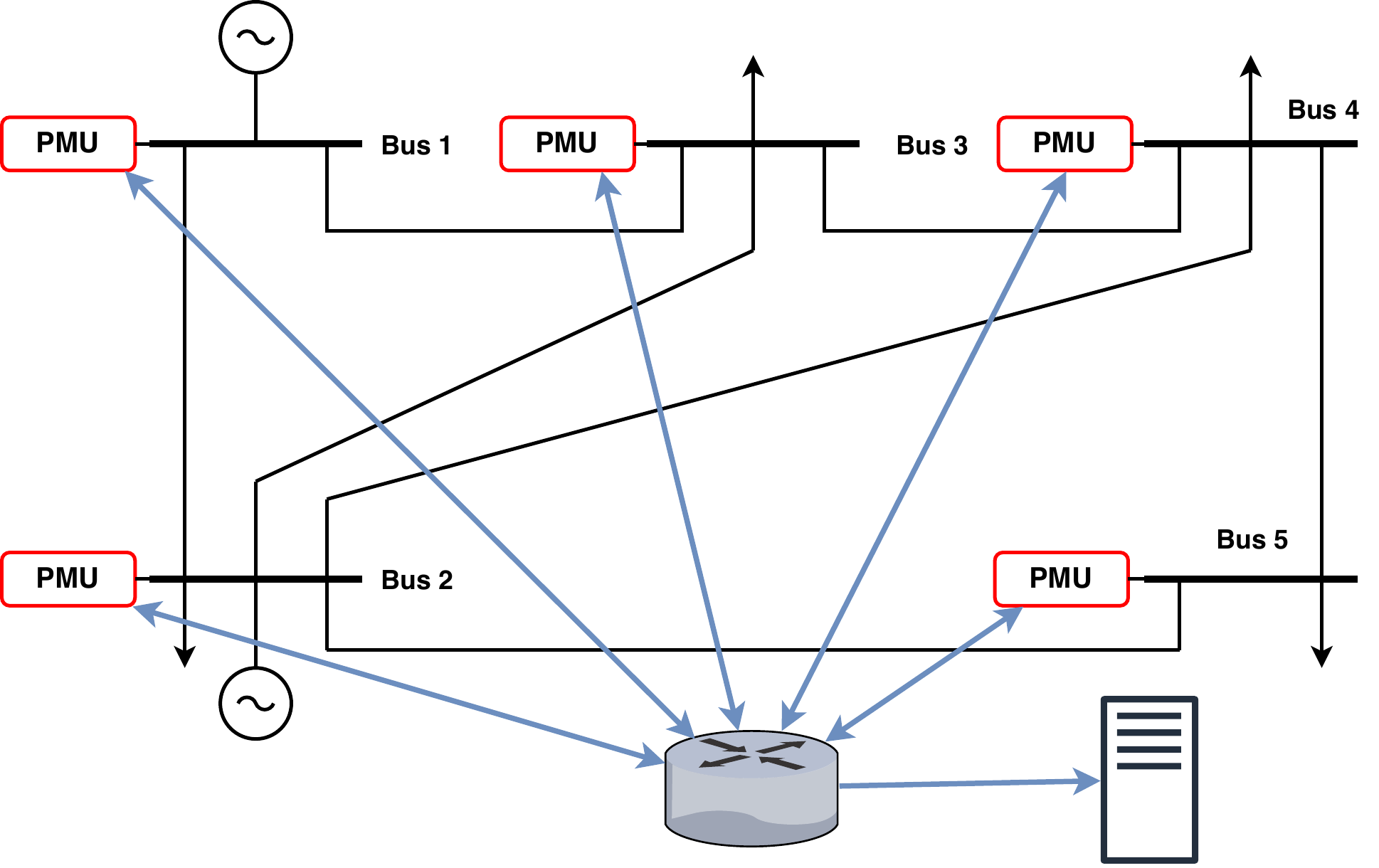}
		\caption{IEEE 5 Bus System} 
		\label{fig:IEEE5bus}
	\end{subfigure}%
	\caption{The experimental setup for testing the framework.} 
	\label{fig:ExpSetup}
	\vspace{-0.4cm}
\end{figure}

The proposed framework is prototyped and tested on a cluster of five Parallella development boards connected in a star network \cite{olo2014ieee} as shown in Fig.\ref{fig:Parallellas}. Parallella contains a Xilinx Zynq SoC (ARM+FPGA) with Linux-capability and a 16-core co-processor which can perform parallel processing. In this work, each IED is a Parallella which would record voltage and current data, gather data from other IEDs, and perform SE if elected as a leader. The framework for communication between agents is created using serf \cite{serf}, which allows broadcast and unicast communication. It provides a platform for devices to execute the challenge-response protocol, and gives them the capability to broadcast events and trigger responses. Each Parallella 32 GB SD card is loaded with voltage and current data generated using MATLAB, representing PMU data (IED performs PMU operation) at each bus of the IEEE 5-bus system shown in Fig.\ref{fig:IEEE5bus}. A server shown in the figure is used to gather data for activity tracking and display purposes. Kalman filter based SE using PMU data \cite{Tebianian2013} is used as the critical algorithm to be secured. Additionally, the scalability of the proposed framework is tested using simulations on the IEEE 118 bus system.


Initially, after the devices are started-up, they are synchronized using the NTP. Then, they start broadcasting information to each other, so that each device can have a view of the network, which is designed using serf. Once they become part of the serf cluster, a leader will be elected, as explained in \cref{subsection:LE}. Now, devices start sending data to the leader periodically (once a minute). \\
\begin{figure}[!htb]
    \begin{subfigure}[t]{0.45\linewidth}
		\centering 
		\captionsetup{justification=centering}
		\includegraphics[width=\linewidth]{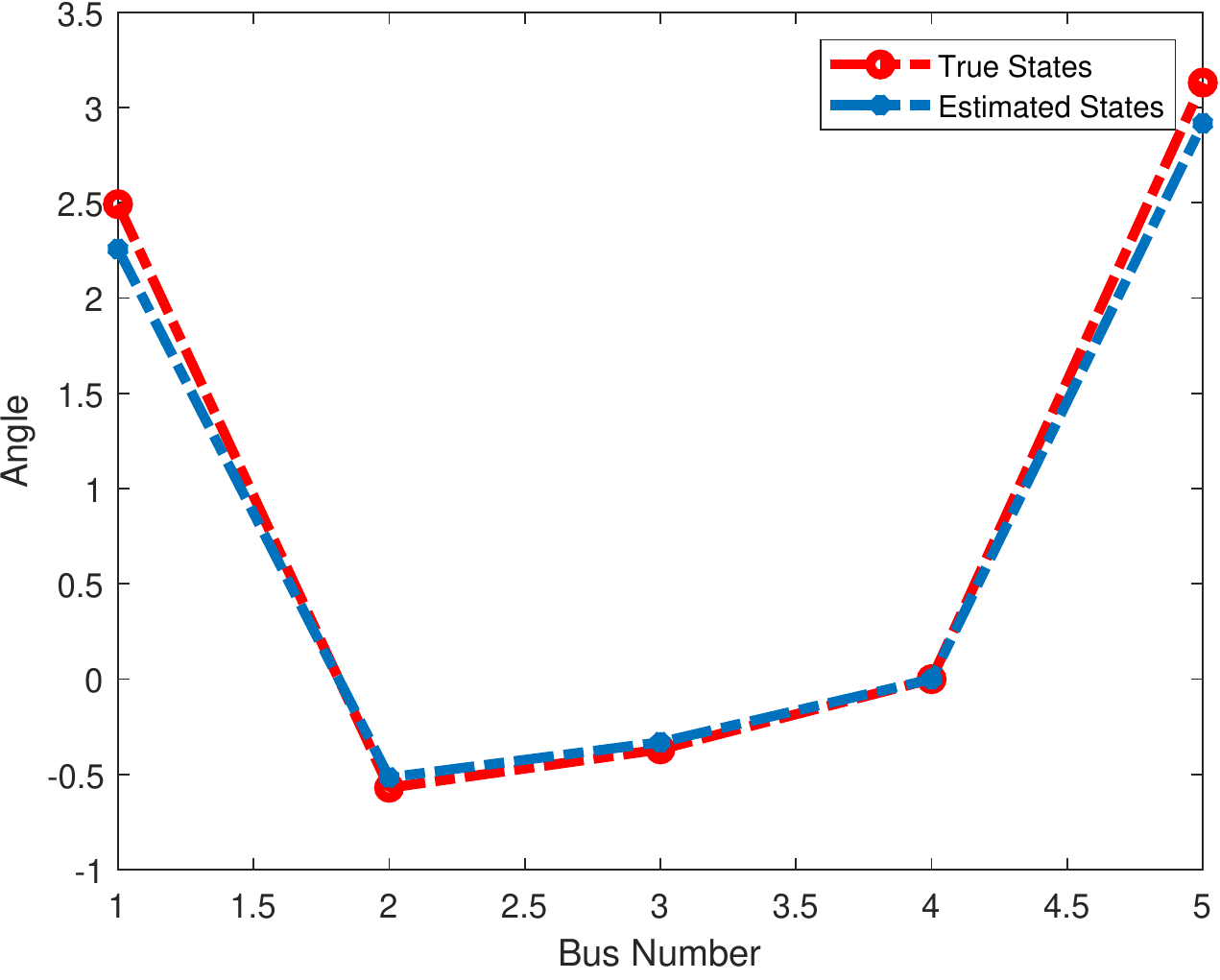}
		\caption{SE, No malicious agent.} 
		\label{fig:SE_noMN}
	\end{subfigure}%
	\begin{subfigure}[t]{0.45\linewidth}
		\centering 
		\includegraphics[width=\linewidth]{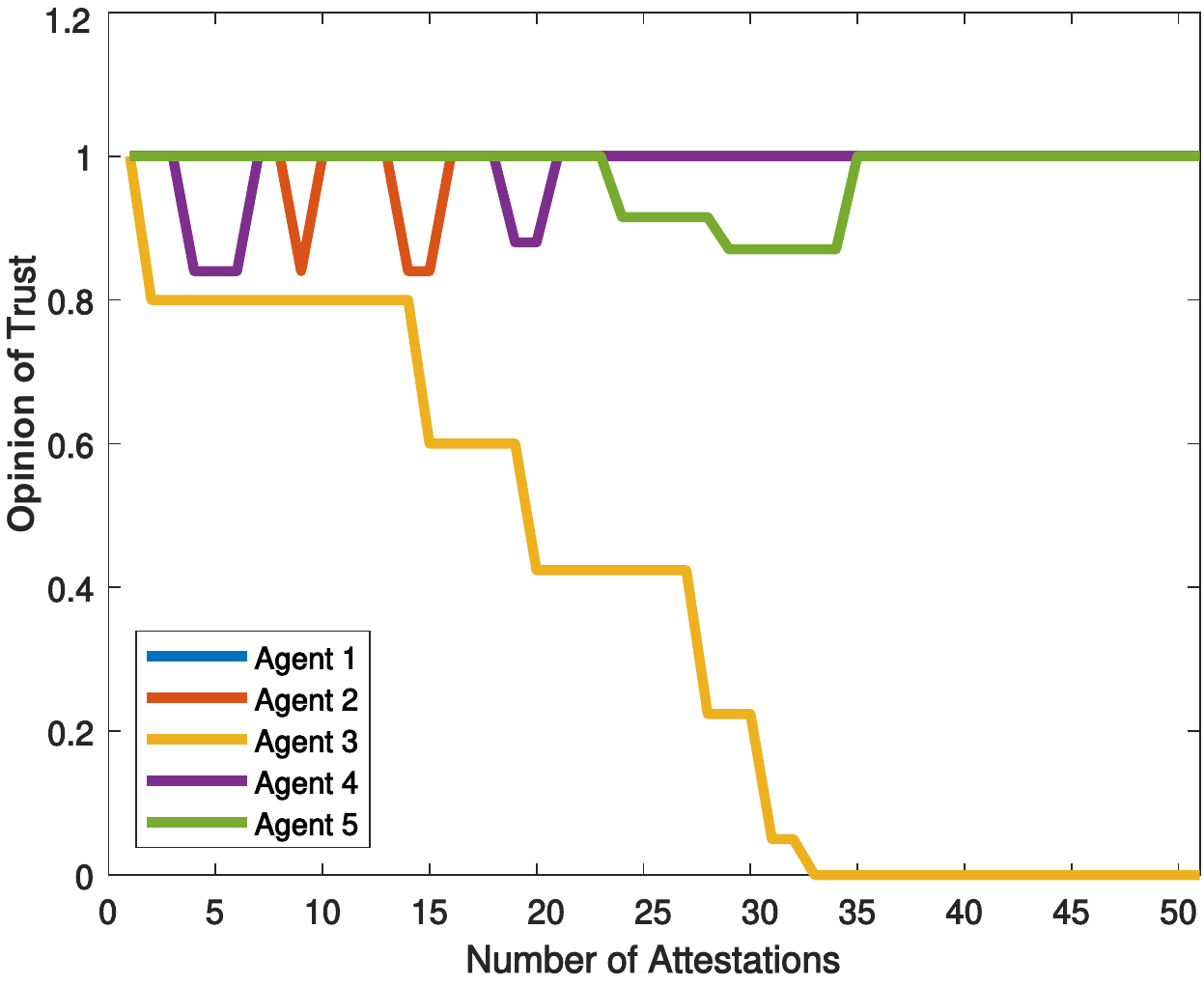}
		\caption{Trust at agent 1.} 
		\label{fig:fig_agent_1}
	\end{subfigure}%
\newline
	\begin{subfigure}[t]{0.45\linewidth}
		\centering 
		\includegraphics[width=\linewidth]{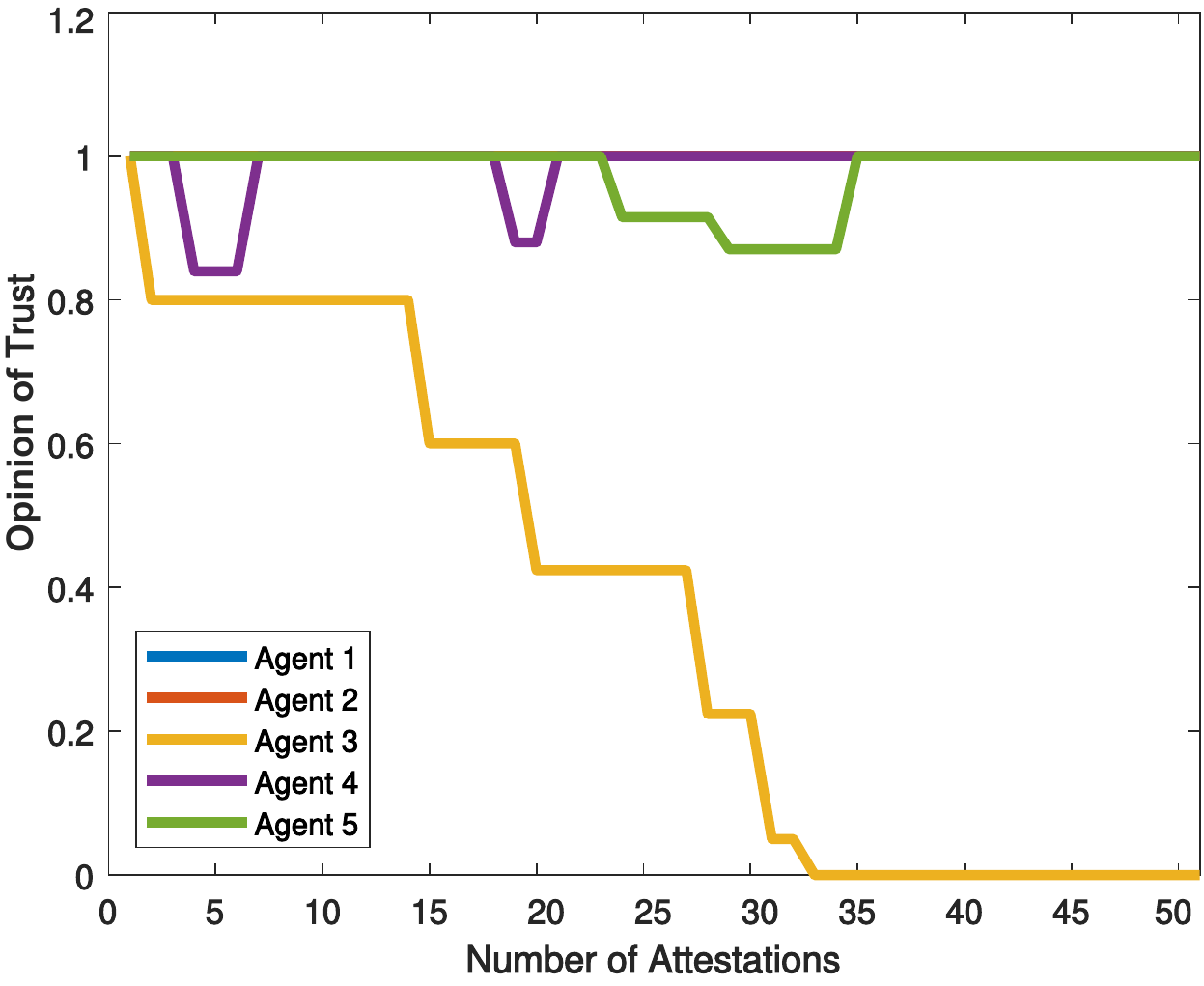}
		\caption{Trust at agent 2.} 
		\label{fig:fig_agent_2}
	\end{subfigure}%
	\begin{subfigure}[t]{0.45\linewidth}
		\centering 
		\includegraphics[width=\linewidth]{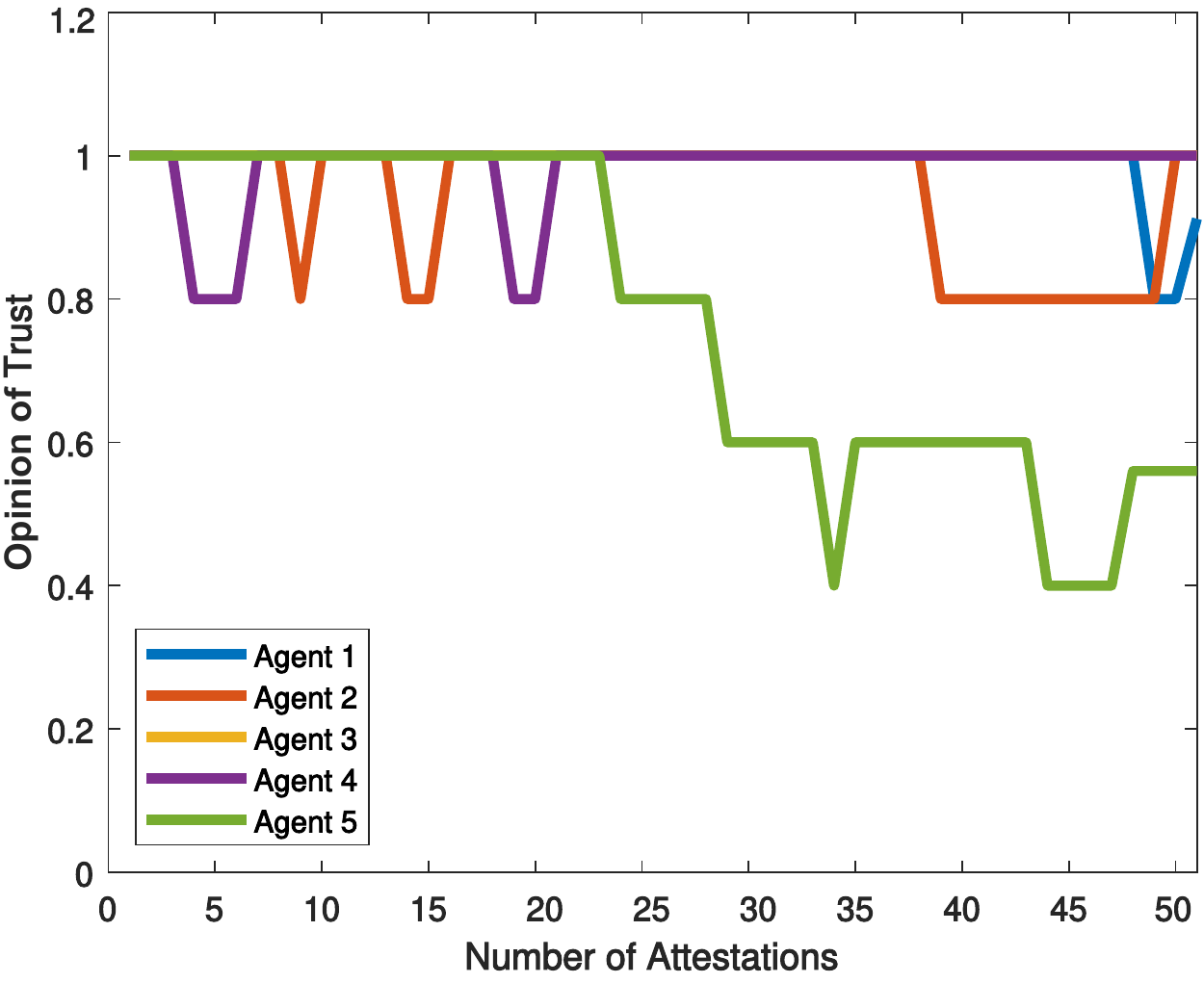}
		\caption{Trust at agent 3.} 
		\label{fig:fig_agent_3}
	\end{subfigure}%
	\newline
	\begin{subfigure}[t]{0.45\linewidth}
		\centering 
		\includegraphics[width=\linewidth]{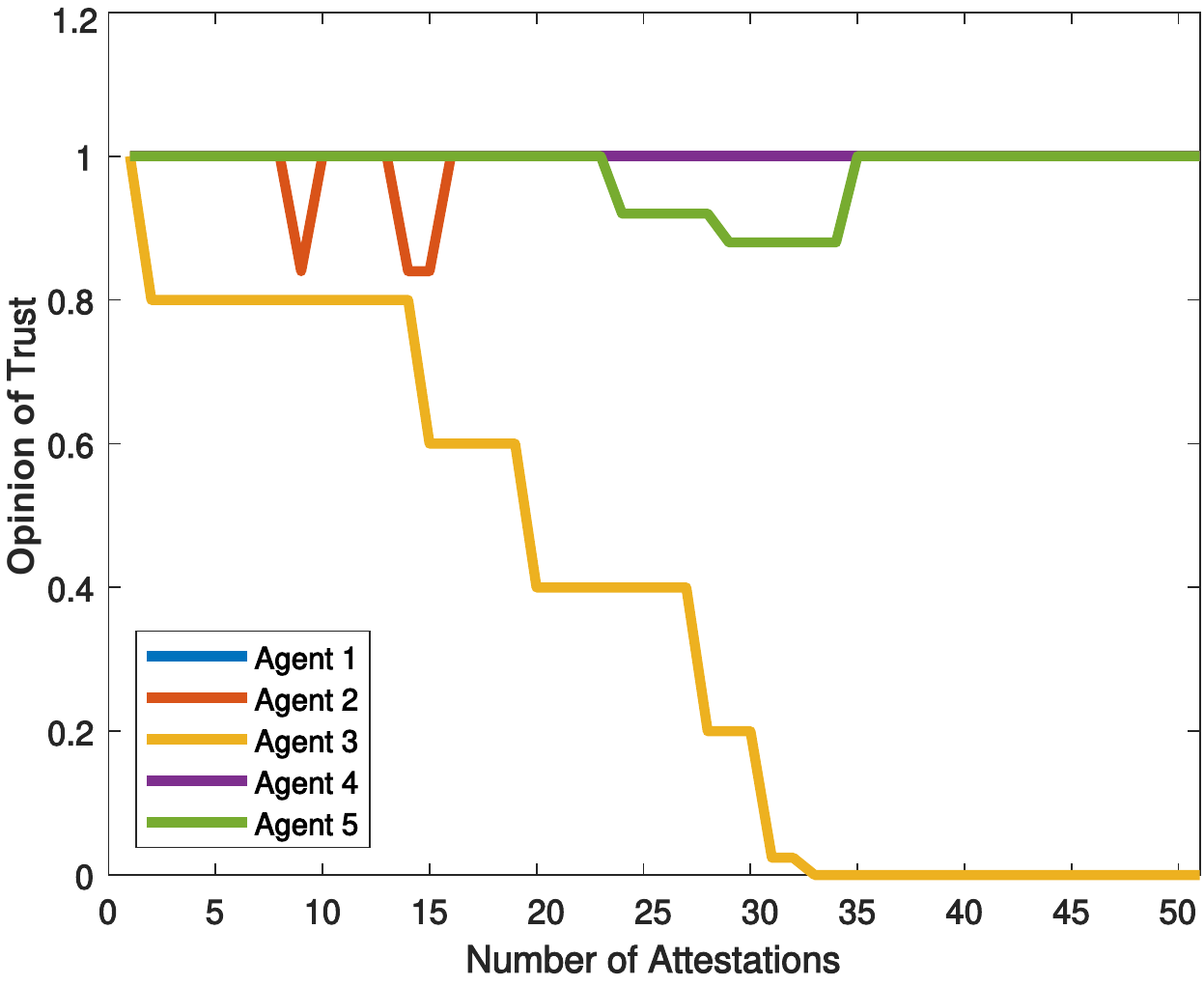}
		\caption{Trust at agent 4.} 
		\label{fig:fig_agent_4}
	\end{subfigure}%
	\centering
	\begin{subfigure}[t]{0.45\linewidth}
		\centering 
		\includegraphics[width=\linewidth]{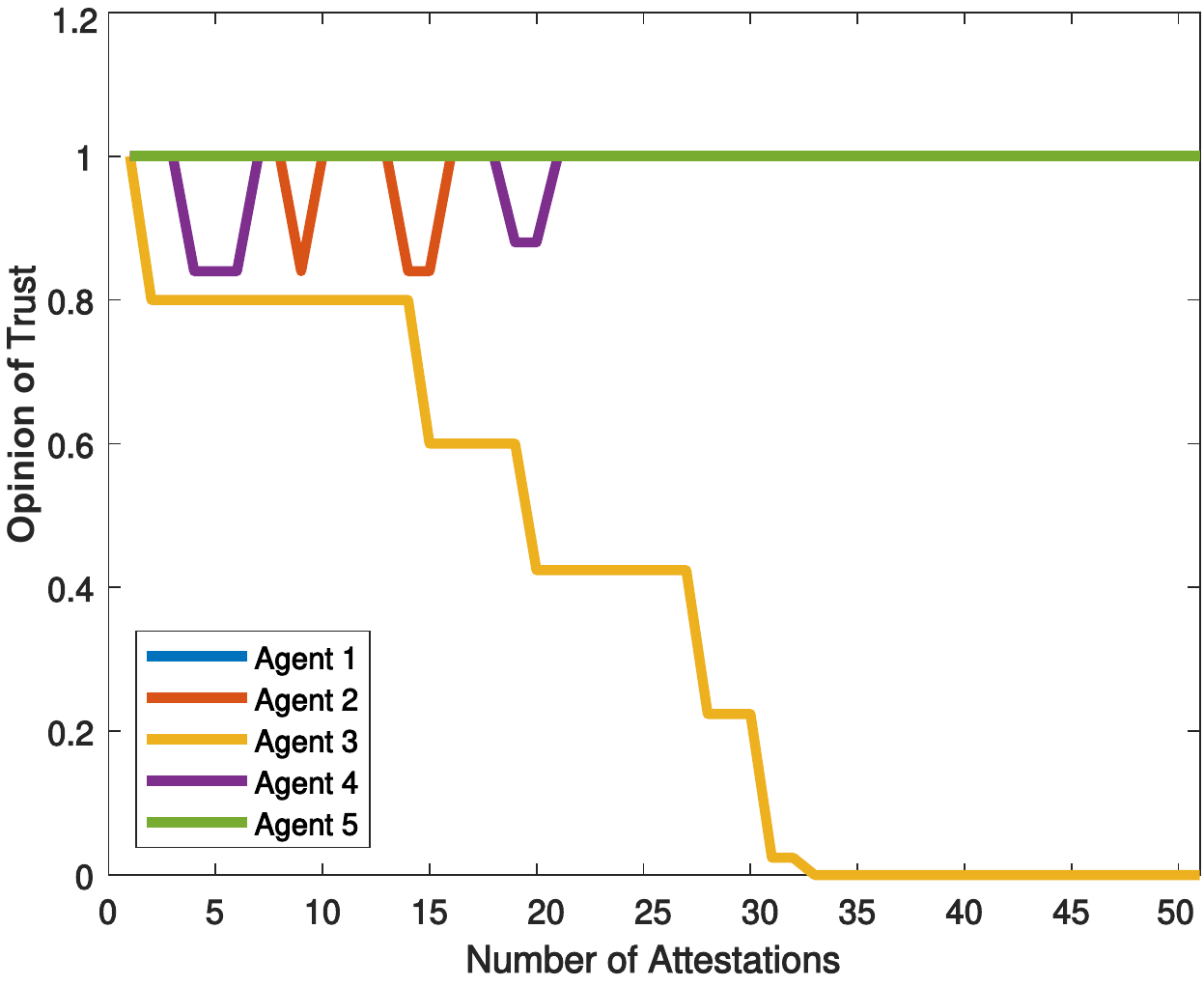}
		\caption{Trust at agent 5.} 
		\label{fig:fig_agent_5}
	\end{subfigure}%
\newline
	\begin{subfigure}[t]{0.45\linewidth}
		\centering 
		\captionsetup{justification=centering}
		\includegraphics[width=\linewidth]{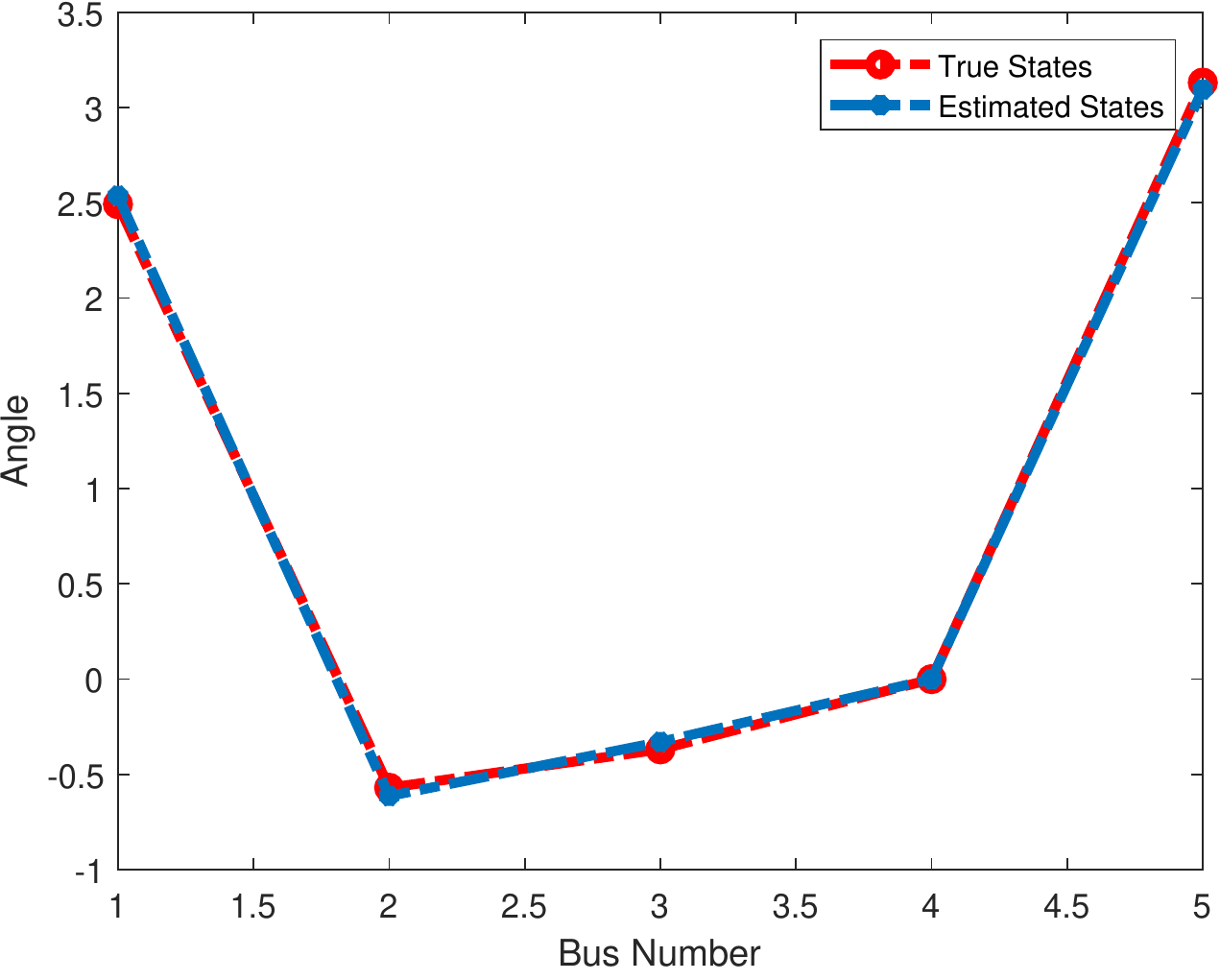}
		\caption{SE, one malicious agent.} 
		\label{fig:SE_MN}
	\end{subfigure}%
	\begin{subfigure}[t]{0.45\linewidth}
		\centering 
		\captionsetup{justification=centering}
		\includegraphics[width=\linewidth]{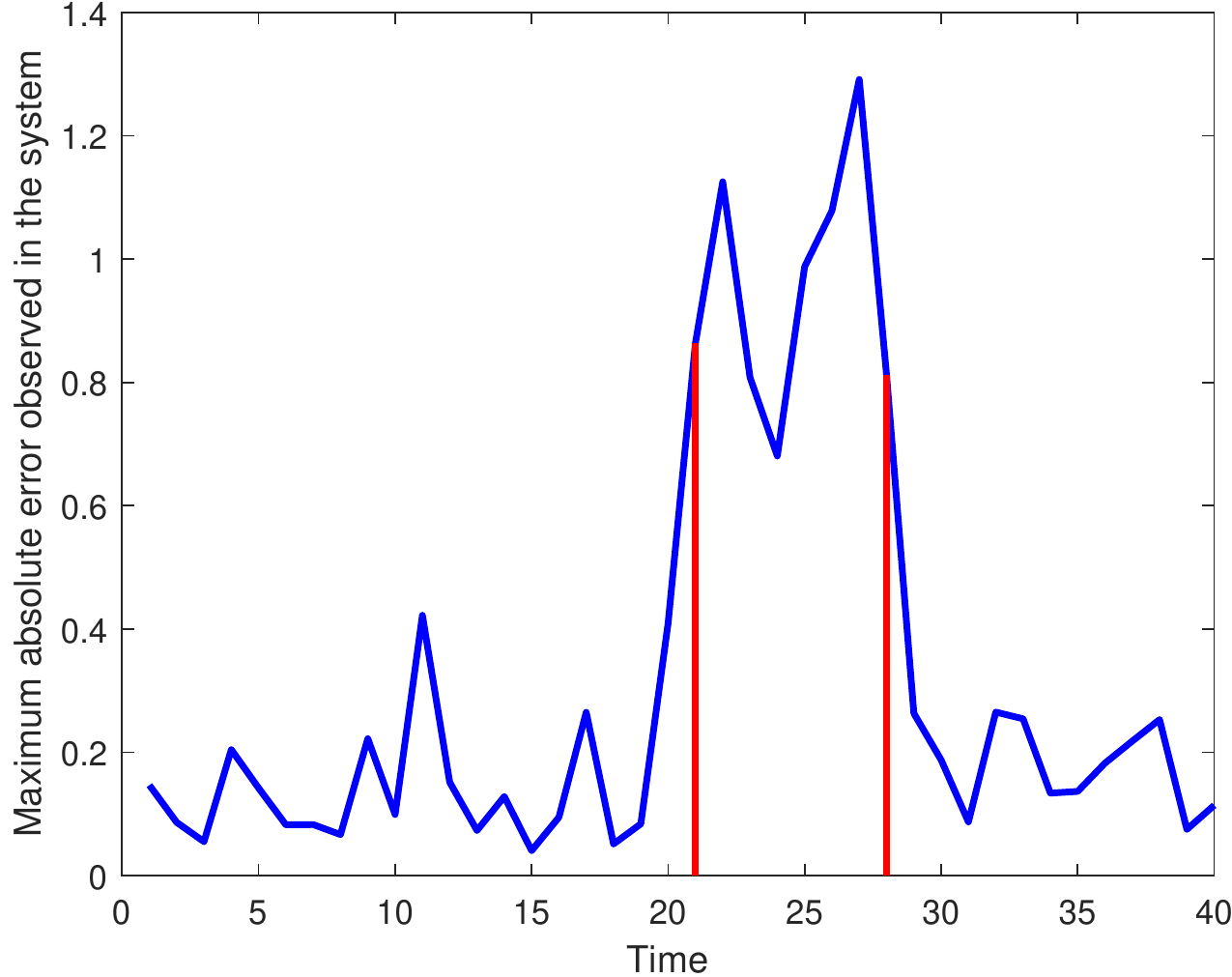}
		\caption{Maximum absolute error.} 
		\label{fig:ErrVsTime}
	\end{subfigure}%
	\caption{5 bus system, with and without a malicious agent.} 
	\label{fig:trust_plots}
	\vspace{-0.9cm}
\end{figure}
\textit{\textbf{SE when there is no malicious agent:}}
The 5 bus system data is taken from the MATPOWER toolbox. SE will begin immediately after the leader election scheme. The estimated phase angles are shown in Fig.\ref{fig:SE_noMN} without any malicious agent. In Fig.\ref{fig:SE_noMN}, the red line indicates true state of the system, whereas the blue line indicates the estimated states. This result is obtained after a single Kalman iteration, and the squared error ($L_2$-norm of the error vector) observed is 0.147.\\
\textit{\textbf{SE when agent at Bus 3 is malicious:}}
The IED at bus 3 is deliberately made malicious by modifying its core process. After the leader election, attestation between devices begins and the trust of a device is updated based on the reports obtained from attestation. In this implementation, every device randomly performs an attestation in a 40 second interval. This interval can be further reduced based on the computational capability of agents. Whenever the trust values for a device drops to $0$ at a majority of devices, data from that device is ignored. The combined choice of attestation and trust management used in the framework is able to detect the malicious device in near real time. The Fig.\ref{fig:trust_plots} shows the evolution of trust at each agent about all the other agents, using \eqref{eqn:trust_model}. One can observe that opinion of trust for agent 3 is reduced to $0$ at agents 1, 2, 4, and 5, as shown in  \Cref{fig:fig_agent_1,fig:fig_agent_2,fig:fig_agent_4,fig:fig_agent_5}, respectively. However its own opinion of trust, in Fig.\ref{fig:fig_agent_3} is very high\textemdash because it does not want to become a target by reducing its own trust. Now, through majority voting, agent 3 will be identified as malicious and its measurements will be subsequently ignored by the leader. Also, it is evident from  \Cref{fig:fig_agent_1,fig:fig_agent_2,fig:fig_agent_3,fig:fig_agent_4,fig:fig_agent_5} that the trust for the other agents is varying, since agent 3 is constantly accusing others of being malicious through negative attestations. However, their trust increases and reaches $1$ due to positive attestations from the other non-malicious agents. This process is constantly running in the background to track any suspicious activity for safe operation of the grid.

The state estimation result of this case is presented in Fig.\ref{fig:SE_MN}. The squared error observed in this case\textemdash after a single Kalman iteration\textemdash is 0.2034. During the simulation, data from device 3 is ignored. Fig.\ref{fig:ErrVsTime}, shows the maximum absolute error observed in the system when a malicious agent is present. The simulation is carried out for a duration of 40 samples. A malicious agent is introduced at the 21\textsuperscript{st} sample, and is detected at the 28\textsuperscript{th} sample; note that the observed error in the system is elevated because of the fabricated data reported by the malicious agent. Once detected, the data from malicious agent is automatically ignored by the leader.

\begin{figure}[!htb]
	\vspace{-0.2cm}
	\begin{subfigure}[t]{0.45\linewidth}
		\captionsetup{justification=centering}
		\includegraphics[width=\linewidth]{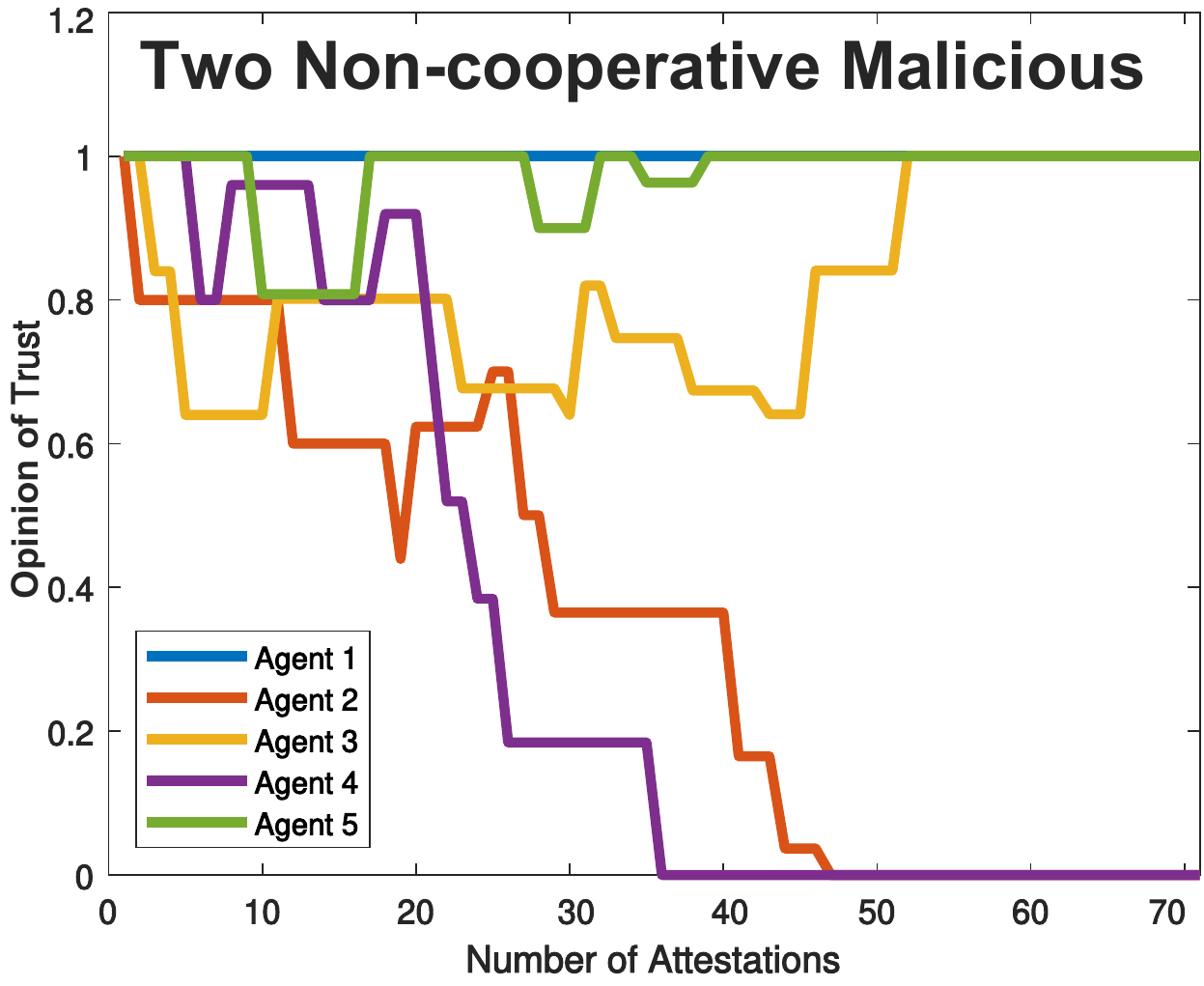}
		\caption{IEEE 5 bus system.} 
		\label{fig:5_NM_trust}
	\end{subfigure}%
	\begin{subfigure}[t]{0.45\linewidth}
		\captionsetup{justification=centering}
		\includegraphics[width=\linewidth]{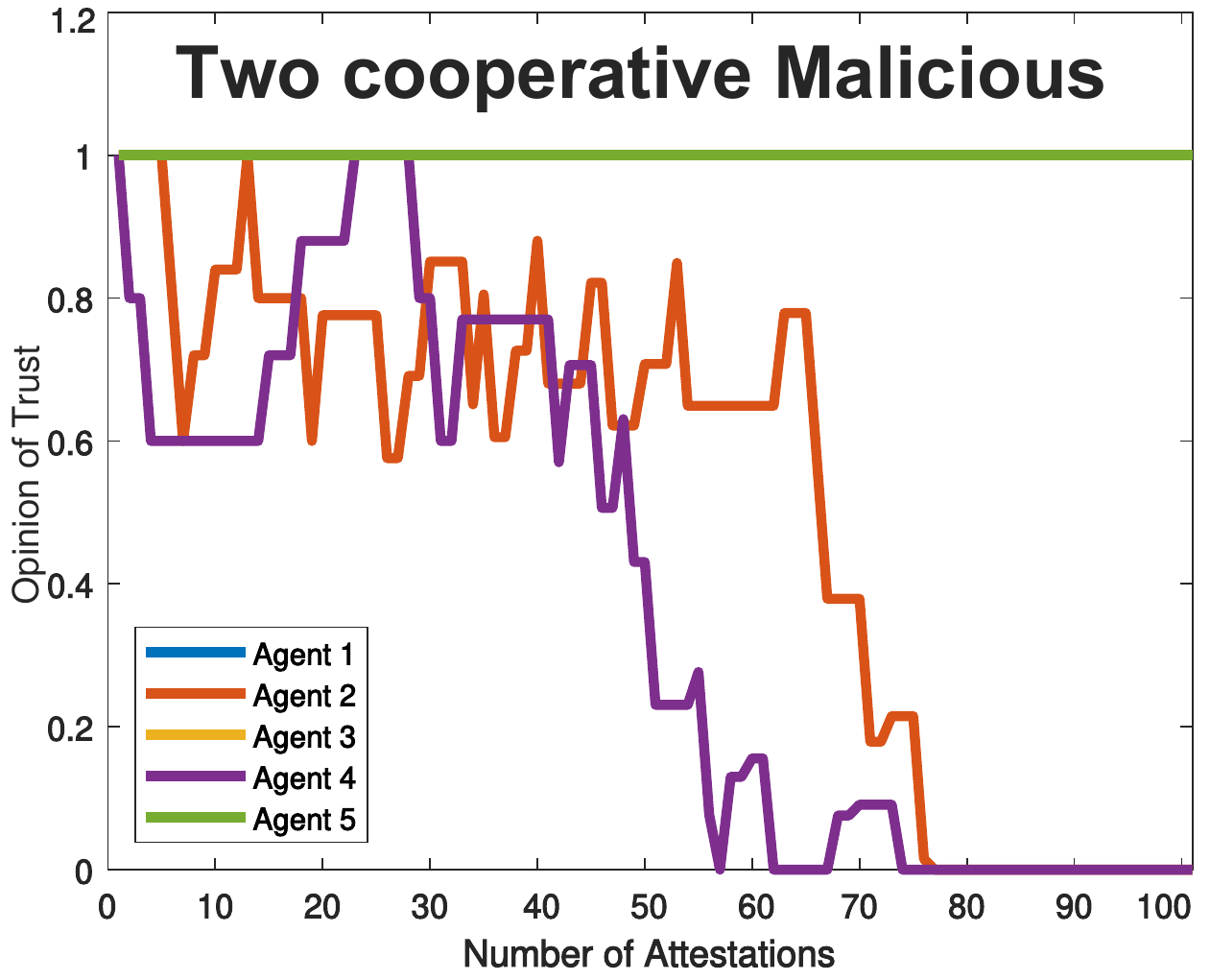}
		\caption{IEEE 5 bus system.} 
		\label{fig:5_NM_trust_Coop}
	\end{subfigure}%
	\label{fig:SE_plots}
	\vspace{0.2cm}
	\newpage
	\begin{subfigure}[t]{0.45\linewidth}
		\centering 
		\captionsetup{justification=centering}
		\includegraphics[width=\linewidth]{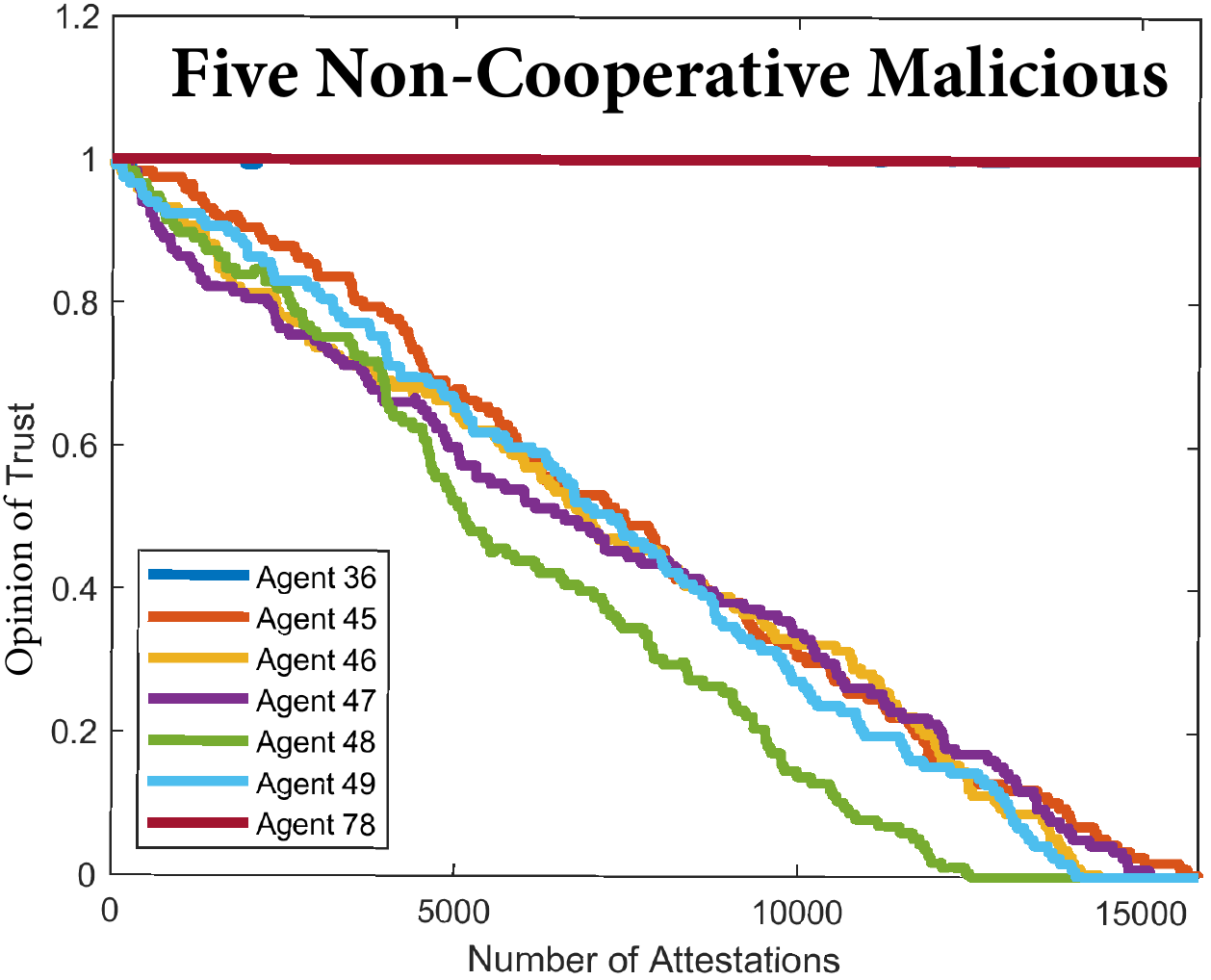}
		\caption{IEEE 118 Bus System.} 
		\label{fig:118_NM_trust}
	\end{subfigure}%
	\begin{subfigure}[t]{0.45\linewidth}
		\centering 
		\captionsetup{justification=centering}
		\includegraphics[width=\linewidth]{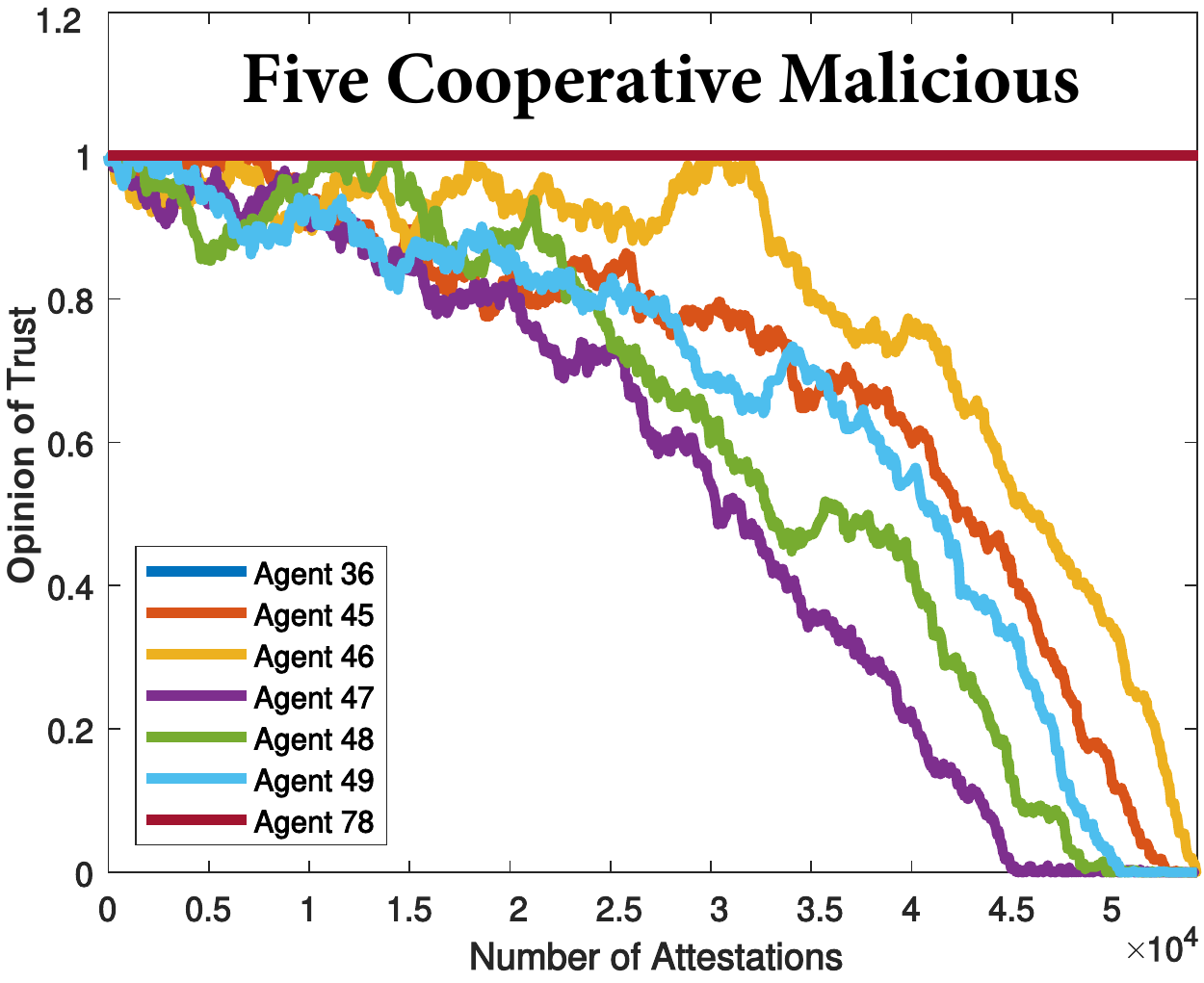}
		\caption{IEEE 118 Bus System.} 
		\label{fig:118_NM_trust_coop}
	\end{subfigure}%
	\caption{Non-malicious agent's trust with $>1$ malicious agents.} 
		\vspace{-0.5cm}
\end{figure}

\textit{\textbf{Multiple malicious agents:}}\label{sec:MMA}
In this section effectiveness of the proposed framework with multiple malicious agents is verified. Two cases, cooperative malicious agents when agents cooperate with each other to increase their own trust and non-cooperative agents when malicious agents do not necessarily cooperate with each other, are evaluated. In the cooperative case, a malicious agent always chooses a malicious agent as an attester whenever it gets a chance of being a verifier and always reports in favor of a malicious agent. In addition to this, when a non-malicious agent reports against one of the malicious agent, all malicious agents increase the trust of this agent instead of reducing it.\\
\textit{\textbf{IEEE 5 Bus System:}}
For the IEEE 5 bus system two malicious agents are created. Evolution of trust at non-malicious agent, is presented for non-cooperative agents and cooperative agents in Fig.\ref{fig:5_NM_trust} and Fig.\ref{fig:5_NM_trust_Coop} respectively. In both simulated cases, the proposed framework is able to identify all the malicious agents in the system, even though the number of malicious agents is $\nless N/2-1$. Furthermore, from Fig.\ref{fig:5_NM_trust} and Fig.\ref{fig:5_NM_trust_Coop}, one can observe that the detection of malicious agents, when they are cooperating with each other, takes more number of attestations in comparison to non-cooperating case.\\
\textit{\textbf{IEEE 118 Bus System:}}
The IEEE 118 bus system is used to test the scalability of the proposed framework and a group of $5$ malicious agents at buses 45 to 49 which represents a big interconnection point with two generator buses, is simulated. Considering the high cost of Parallella devices, we resort to simulations on this test system to evaluate the proposed framework. Simulation results on 118 bus systems for both non-cooperative and cooperative cases are presented in Fig.\ref{fig:118_NM_trust} and Fig.\ref{fig:118_NM_trust_coop}, respectively. Though the framework is able to identify all the malicious agents, the number of attestations performed in both the cases are high. For a bigger system with hundreds of buses, the proposed framework might take some time to reach a consensus through majority voting. Therefore, suitable methods to reduce the number of attestations can be explored as one of the future research directions. One approach could be that we sub-divide the network into different areas and a few nodes in an area can be designated for leader election in each cluster; this reduces the number of attestations and costs associated with communication infrastructure.

\begin{figure}[!htb]
\begin{subfigure}[t]{0.5\linewidth}
	\centering 
	\captionsetup{justification=centering}
	\includegraphics[width=\linewidth,height = 1in]{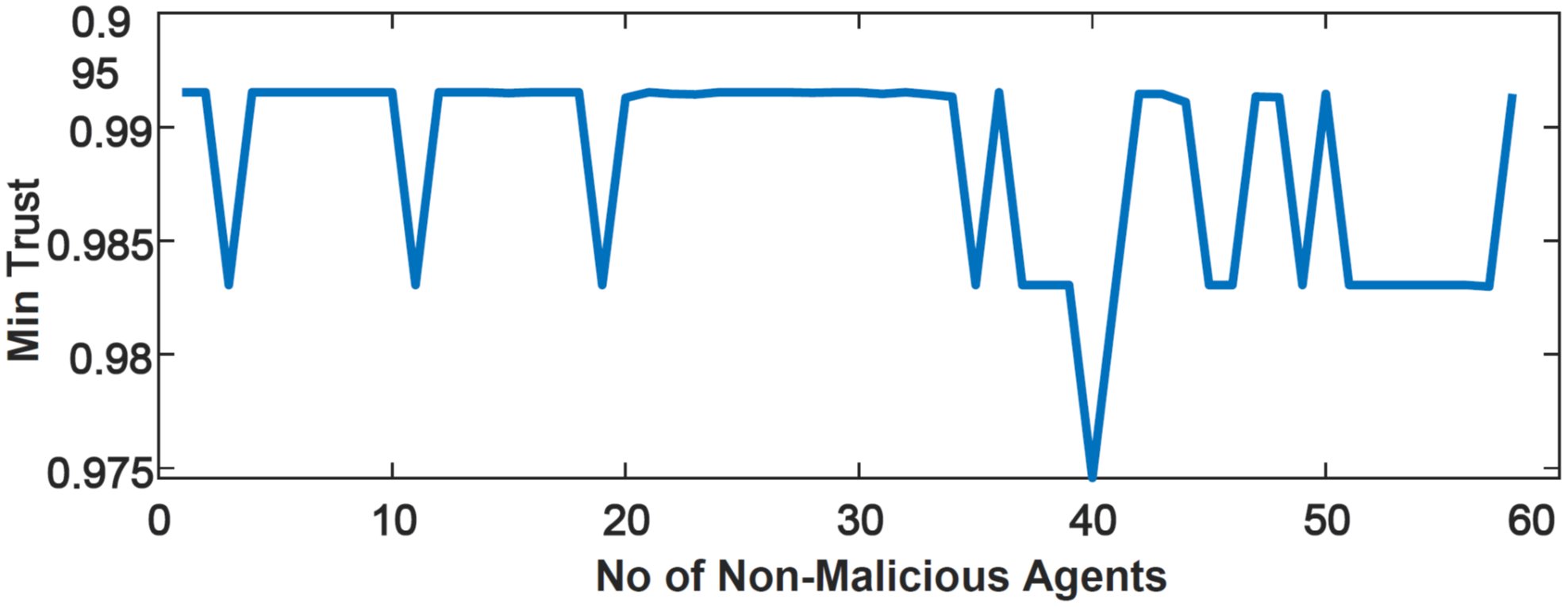}
	\caption{Min trust, non-mal. agents.} 
	\label{fig:minvsmax}
\end{subfigure}%
\begin{subfigure}[t]{0.5\linewidth}
	\centering 
	\captionsetup{justification=centering}
	\includegraphics[width=\linewidth, height = 1in]{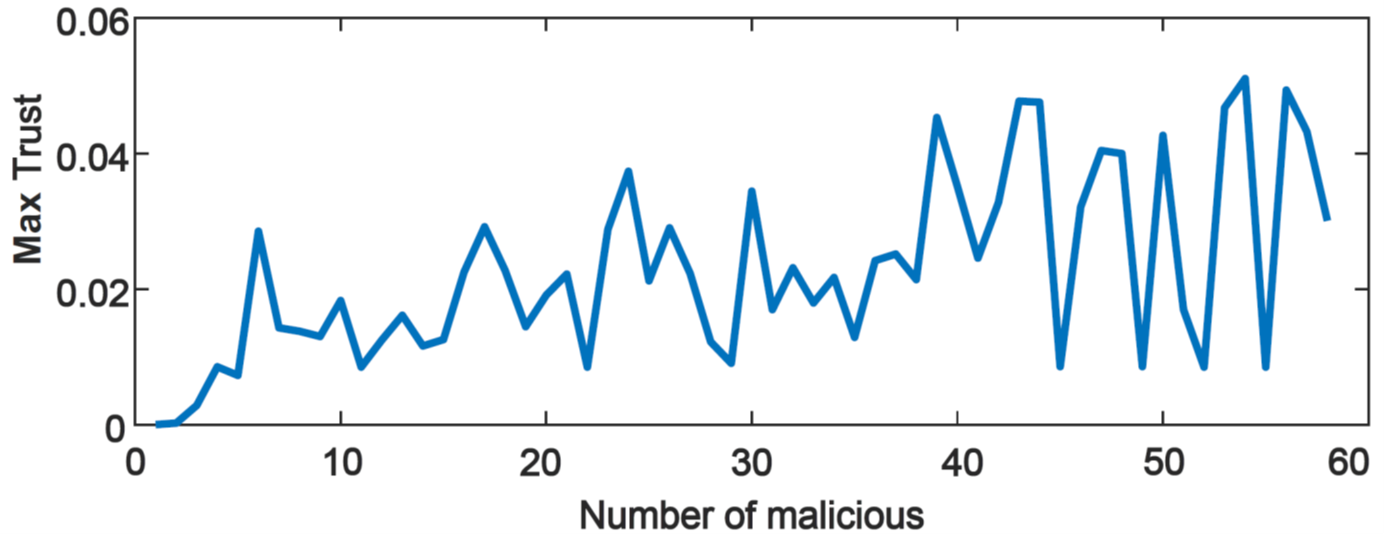}
	\caption{Max trust, malicious agents} 
	\label{fig:118bus_trust}
\end{subfigure}%
\caption{Min. and max. values of trust, 118 bus system.}
\vspace{-0.4cm}
\end{figure}

The proposed framework will be able to handle any number of malicious agents that is strictly less than $\frac{N}{2}-1$; in this case the ordinary differential equation starting from the all ones (trusting) initial condition settles at $p^*$ where all malicious and honest agents are correctly identified. It can be verified empirically as follows. In Fig.\ref{fig:minvsmax}, the minimum and maximum values of trust for non-malicious and malicious agents are presented. The simulation is stopped the moment $\frac{N}{2}$ or more honest agents identify the status of each agent correctly. The first plot in Fig.\ref{fig:minvsmax} represents the minimum value of trust for the non-malicious agents, and the second plot represents the maximum value of trust for the malicious agents. From Fig.\ref{fig:minvsmax}, one can observe that the worst-case non-malicious agents minimum trust values always exceed the maximum of the malicious agents trust values, i.e., the proposed framework is able to identify all the malicious agents (via a majority rule) as long as they are strictly less than $\frac{N}{2}-1$ in number.

\vspace{-0.4cm}
\section{Conclusion}
In this paper a distributed hierarchy-based framework to protect critical functions of a coordinating control center in a distributed environment is proposed. Kalman filter based SE is used as the critical function that is to be secured for demonstration. The protection comes from choosing the agent that becomes the coordinator in a random fashion, from an attestation and trust management protocol that maintains and updates the trust metric of every agent in the system, and from a consensus algorithm that isolates the malicious agents. All these are distributed algorithms. Assertions coming from attestations are used to measure trust. A graceful trust update protocol ensures that the malicious agents' trust have a tendency to decrease if there are enough honest agents deemed as honest. The malicious agents are eventually identified as malicious as empirically verified in the simulations. The theoretical results identifies the points of settlement of the trust management algorithm. While there is a nonzero probability that the trust values may settle at undesirable values, if the number of malicious agents is strictly less than $\frac{N}{2}-1$ and the initial trust values are all 1, then the system has a high probability of settling at the values where all agents statuses are correctly identified by all the honest agents, as verified empirically in the simulations and experimental setup. The design results in a logical hierarchy from the perspective of the critical functionality (SE), since it is performed at a randomly elected leader/coordinator agent. We have thus demonstrated a decentralized framework to protect a critical function (such as SE) at a control center. 

\vspace{-0.4cm}
\appendices
\section{Proof of Proposition~\ref{prop:main}}
\label{appendix:proof-of-proposition-main}
\begin{scriptsize}
Let us recall the iteration
\vspace{-0.2cm}
\begin{eqnarray*}
  p_{ij}(t+1) = \left[ p_{ij}(t) + \Delta_{k \rightarrow j}(p_{ik}(t)) \right]_0^1, \quad
  \forall i \in \mathcal{H}, 1 \leq j \leq N, j \neq i, j \neq k
\end{eqnarray*}
when agent $k$ attests agent $j$ at time $t+1$. The notation. $[\cdot]_0^1$ refers to the projection operation.  If $i=j$, take $p_{ii}(t) = 1$. Recall that $e_j = 1$ if $j \in \mathcal{H}$ and $e_j=-1$ otherwise. Write ${\bf 1}\{\cdot\}$ for the indicator function of an event. Recognising that the choice of the agent pair $e(t+1)$ for the verifier and attester is uniformly random across the agents
\begin{small}
\begin{align*}
  & p_{ij}(t+1) = \Big[ p_{ij}(t) + \sum_{k: k \neq j} \Delta_{k \rightarrow j}(p_{ik}(t)) {\bf 1}\{ e(t+1) = (k,j) \} \Big]_0^1 \\
  & \stackrel{(\ref{eqn:indicator2})}{=} \Big[ p_{ij}(t) + a(t) \sum_{k: k \neq j} e_k e_j p_{ik}(t) {\bf 1}\{ e(t+1) = (k,j) \} \Big]_0^1 \\
  & = \Big[ p_{ij}(t) + a(t) \Big( \frac{e_j}{N(N-1)} \sum_{k : k \neq j} e_k p_{ik}(t) \\
  & + \sum_{k: k \neq j} e_k e_j p_{ik}(t) \left( {\bf 1} \{ e(t+1) = (k,j) \} - \frac{1}{N(N-1)} \right) \Big) \Big]_0^1 \\
  & = \Big[ p_{ij}(t) + a(t) \Big( \frac{e_j}{N(N-1)} (p_{iH}^{(j)}(t) - p_{iM}^{(j)}(t))+ M_{ij}(t+1) \Big) \Big]_0^1
\end{align*}\end{small} 
where $M_{ij}$ is a martingale noise (with respect to the filtration generated by the $p(\cdot) = (p_{ij}(\cdot), i \in \mathcal{H}, 1 \leq j \leq N)$ process. If the projection operation were not present, then one would have anticipated that the iterates would track the ordinary differential equation system
\begin{equation}
  \label{eqn:false-dynamics}
  \dot{p}_{ij}(t) = \frac{e_j}{N(N-1)} (p_{iH}^{(j)}(t) - p_{iM}^{(j)}(t)).
\end{equation}
However, the projection operation changes the dynamics, and the goal is to identify this modified dynamics. 

Let us write $p := (p_{ij}, i \in \mathcal{H}, 1 \leq j \leq N)$ and $[p]_0^1 := ([p_{ij}]_0^1, i \in \mathcal{H}, 1 \leq j \leq N)$. The projection operation is defined component-wise, and so the Fr\'{e}chet derivative of $[p]$ at $p$ in the direction $q$, denoted $\gamma(p;q)$ and defined by
\[
  \gamma(p; q) := \lim_{\delta \downarrow 0} \frac{[p + \delta q]_0^1 - p}{\delta},
\]
is given by \vspace{-0.5cm}
\begin{equation}
  \label{eqn:Frechet}
  \gamma(p;q)_{ij} = 
  \begin{cases}
    0 & p_{ij} = 1 \mbox{ and } q_{ij} > 0 \\
    0 & p_{ij} = -1 \mbox{ and } q_{ij} < 0 \\
    q_{ij} & \mbox{otherwise}.
  \end{cases}
\end{equation}
Since the martingale noise vector, whose $ij$ component is
\begin{small}
\[
M_{ij}(t) = \sum_{k: k \neq j} e_k e_j p_{ik}(t) \left( {\bf 1} \{ e(t+1) = (k,j) \} - \frac{1}{N(N-1)} \right),
\]
\end{small} is conditionally independent of the history given $p(t)$, the driving vector field for the dynamics is not given by the right-hand side (\ref{eqn:false-dynamics}), but instead is given  by
\begin{equation}
    \label{eqn:true-dynamics}
    h(p(t)) := E \left[  \gamma \left(p(t) ; q(t) \right) ~|~ p(t) \right]
\end{equation}
where
\begin{equation}
  \label{eqn:tangent}
    \begin{array}{l}
  q(t) = \sum_{k: k \neq j} e_k e_j p_{ik}(t) {\bf 1}\{ e(t+1) = (k,j), \}, ~i \in \mathcal{H}, 1 \leq j \leq N.
  \end{array} 
  \end{equation}
Using (\ref{eqn:Frechet}) and (\ref{eqn:tangent}) in (\ref{eqn:true-dynamics}), one gets
\begin{eqnarray*}
  h(p(t))_{ij} & = & \frac{1}{N(N-1)} \Big[
  e_j(p_{iH}(t) - p_{iM}(t)) \cdot {\bf 1}\{ p_{ij}(t) \in (0,1)\} \\
  & & \hspace*{1.8cm} - \left[ e_j(p_{iH}(t) - p_{iM}(t)) \right]_{-} \cdot {\bf 1}\{ p_{ij}(t) = 1\} \\
  & & \hspace*{1.8cm} + [e_j(p_{iH}(t) - p_{iM}(t))]_{+} \cdot {\bf 1}\{ p_{ij}(t) = 0\}
  \Big]
\end{eqnarray*}
which is the same as the driving function (\ref{eqn:drivingfunction}). It is easy to check that the driving function mapping $p \mapsto h(p)$ is not continuous at some boundary points. Then, by the results of \cite[Sec.~5.4]{borkar2009stochastic}, all subsequential limits of the iterates track the differential inclusion $\dot{p}(t) \in h_E(p(t))$ and, further, the iterates converge to a possibly random, connected, compact, internally chain transitive, invariant set for the differential inclusion $\dot{p}(t) \in h_E(p(t))$. This completes the proof. \IEEEQEDclosed
\end{scriptsize}
\vspace{-0.3cm}

\ifCLASSOPTIONcaptionsoff 
  \newpage
\fi

\bibliographystyle{IEEEtran}
\bibliography{IEEEfull}

\end{document}